\documentclass[12pt]{article}

\usepackage[superscript,sort]{cite}

\usepackage{ifthen,ifpdf}
\ifpdf
	\usepackage[pdftex]{hyperref}	\hypersetup{colorlinks=true,linkcolor=black,citecolor=black,urlcolor=blue,backref=page,bookmarks=true,breaklinks=true,plainpages=false }
	\usepackage[pdftex]{graphicx}
	\usepackage[usenames,pdftex]{color}
\else
	\usepackage[colorlinks=true,breaklinks=true,plainpages=false]{hyperref}
	\usepackage{graphicx}
	\usepackage[usenames]{color}
\fi

\usepackage{amsthm,amscd,amsxtra,amsfonts,amsmath,amssymb,multirow}
\usepackage{wrapfig}
\usepackage[footnotesize]{caption}
\usepackage[tiny,compact]{titlesec}
\usepackage[textwidth=0.8in,textsize=footnotesize]{todonotes}
\usepackage{algorithm,algorithmic,extarrows}

\begin{document}


\title{Persistent Homology for The Quantitative Prediction of Fullerene Stability
}

\author{
Kelin Xia$^1$,
Xin Feng$^2$,
Yiying Tong$^2$\footnote{Corresponding author.  E-mail:ytong@msu.edu},
and
Guo Wei Wei$^{1,3}$\footnote{Corresponding author.  E-mail:wei@math.msu.edu}\\
$^1$Department of Mathematics \\
Michigan State University, MI 48824, USA\\
$^2$Department of Computer Science and Engineering \\
Michigan State University, MI 48824, USA\\
$^3$Department of Biochemistry  and Molecular Biology \\
Michigan State University, MI 48824, USA \\
}

\date{\today}
\maketitle

\begin{abstract}
Persistent homology is a relatively new tool often used for \emph{qualitative} analysis of intrinsic topological features in images and data originated from scientific and engineering applications.   In this paper, we report novel \emph{quantitative} predictions of   the energy and stability of fullerene molecules, the very first attempt in employing  persistent homology in this context. The ground-state structures of a series of small fullerene molecules are first investigated  with the standard Vietoris-Rips complex. We decipher all the barcodes, including both short-lived local bars and long-lived global bars arising from topological invariants, and associate them with fullerene structural details. By using  accumulated bar lengths, we build quantitative models to correlate local and global Betti-2 bars respectively with the heat of formation and total curvature energies of fullerenes.  It is found that the heat of formation energy is related to the local hexagonal cavities of small fullerenes, while the total curvature energies of fullerene isomers are associated with their sphericities, which are measured by the lengths of their long-lived Betti-2 bars. Excellent correlation coefficients ($>0.94$) between persistent homology predictions and those of quantum or curvature analysis have been observed.  {A correlation matrix based filtration is introduced to further verify our findings}.
\end{abstract}

{\bf Key words:} persistent homology, filtration, fullerene,  isomer, nanotube, stability, curvature.

\newpage

{\setcounter{tocdepth}{5} \tableofcontents}

\newpage


\section{Introduction}\label{Sec:Introduction}
Persistent homology, a method for studying topological features over changing scales, has received tremendous attention  in the past decade \cite{Edelsbrunner:2002,Zomorodian:2005}. The basic idea is to measure the life cycle of topological features within a filtration, i.e., a nested family of abstract simplicial complexes, such as Vietoris-Rips complexes, \v{C}ech complexes,  or alpha complexes \cite{Edelsbrunner:1994}. Thus, long-lived topological characteristics, which are often  the intrinsic invariants of the underlying system, can be extracted; while short-lived features are filtered out. The essential topological characteristics of three-dimensional (3D) objects typically include connected components, tunnels or rings, and cavities or voids, which are invariant under the non-degenerate deformation of the structure. Homology characterizes such  structures as groups, whose generators can be considered independent components, tunnels, cavities, etc. Their times of ``birth'' and ``death'' can be measured by a function associated with the filtration, calculated with ever more efficient computational procedures \cite{edelsbrunner:2010,Dey:2008,Dey:2013,Mischaikow:2013}, and further visualized through barcodes \cite{Ghrist:2008}, a series of horizontal line segments with the $x$-axis representing the changing scale and the $y$-axis representing the index of the homology generators.  Numerous software packages, such as Perseus, Dionysus, and Javaplex \cite{javaPlex}, based on  various algorithms have been developed and made available in the public domain.
As an efficient tool to  {unveil topological invariants}, persistent homology has been applied to various fields, such as image analysis \cite{Carlsson:2008,Pachauri:2011,Singh:2008}, chaotic dynamics verification \cite{Mischaikow:1999,Kaczynski:2004}, sensor network \cite{Silva:2005}, complex network \cite{LeeH:2012,Horak:2009}, data analysis \cite{Carlsson:2009},  {geometric processing}\cite{Feng:2013}, and computational biology \cite{Kasson:2007,Gameiro:2013,Dabaghian:2012,YaoY:2009}. Based on persistent homology analysis, we have proposed molecular topological fingerprints and utilized them to reveal the topology-function relationship of biomolecules \cite{KLXia:2014c}.  In general, persistent homology is devised as a robust but \emph{qualitative} topological tool and has been hardly employed as a precise \emph{quantitative} predictive  {tool \cite{Adcock:2013,Bendich:2014}.}

To the best of our knowledge, persistent homology has not been applied to the study of fullerenes, special molecules comprised of only carbon atoms. The fullerene family shares the same closed carbon-cage structure, which contains only pentagonal and hexagonal rings. In 1985, Kroto et al. ~\cite{Kroto:1985} proposed the first structure of C$_{60}$, which  {was} then confirmed in 1990 by Kr$\ddot{a}$tschmer et al. \cite{Kratschmer:1990} in synthesizing macroscopic quantities of C$_{60}$. Enormous  interest has been aroused by these interesting discoveries.  However, there are many challenges. Among them, finding the ground-state structure has been a primary target.

In general, two types of approaches are commonly used \cite{Fowler:1988,Manolopoulos:1991,Fowler:1995,Ballone:1990,Chelikowsky:1991,ZhangBL:1992a}. The first method is based on the geometric and topological symmetries of fullerene \cite{Fowler:1988,Manolopoulos:1991,Fowler:1995}. In this approach, one first constructs all possible isomers, and then chooses the best possible candidate based on the analysis of the highest-occupied molecular orbital (HOMO) energy and the lowest-unoccupied molecular orbital (LUMO) energy~\cite{Manolopoulos:1991}. In real applications, to generate all possible isomers for a fullerene with a given atom count is nontrivial until the introduction of Coxeter's construction method ~\cite{Coxeter:1971,Fowler:1988} and the ring spiral method~\cite{Manolopoulos:1991}. In Coxeter's method, the icosahedral triangulations of the sphere are analyzed to evaluate the possible isomer structures. This method is mathematically rigorous. However, practical applications run into issues with low-symmetry structures. On the other hand, based on the  {spiral conjecture\cite{Fowler:1995}}, the ring spiral method simply lists all possible spiral sequences of pentagons and hexagons, and then winds them up into fullerenes. When a consistent structure is found, an isomer is generated; otherwise, the sequence is skipped. Although the conjecture breaks down for fullerenes with 380 or more atoms, the spiral method proves to be quite efficient \cite{Fowler:1995}.

For each isomer, its electronic structure can be modeled simply by the H\"{u}ckel molecular orbital theory \cite{Streitwieser:1961}, which is known to work well for planar aromatic hydrocarbons using standard C-C and C-H $\sigma$ bond energies. Similarly, the bonding connectivities in fullerene structures are used to evaluate orbital energies. The stability of the isomers, according to Manolopoulus \cite{Manolopoulos:1991}, can then be directly related to the calculated HOMO-LUMO energy gap. However, this model falls short for large fullerene molecules. Even for small structures, its prediction tends to be inaccurate. One possible reason is fullerene's special cage structures. Instead of  {a planar shape}, the structure usually has local curvatures, which jeopardizes the $\sigma$-$\pi$ orbital separation \cite{Fowler:1994,Fowler:1995}. To account for curvature contributions, a strain energy is considered. It is found that the stain energy reaches  {its minimum when pentagonal faces} are as far away as possible from each other. This is highly consistent with the isolated pentagon rule (IPR) ---  the most stable fullerenes are those in which all the pentagons are isolated \cite{Fowler:1995}.

Another approach to obtain ground-state structures for fullerene molecules is through simulated annealing \cite{Ballone:1990,Chelikowsky:1991,ZhangBL:1992a}. This global optimization method works well for some structures. However, if large energy barriers exist in the potential, the whole system is prone to be trapped into metastable high-energy state. This happens as breaking the carbon bonds and rearranging the structure need a huge amount of energy. A revised method is to start the system from a special face-dual network and then employ the tight-binding potential model \cite{ZhangBL:1992a,ZhangBL:1992b}. This modified algorithm manages to generate the  C$_{60}$ structure of $I_h$ symmetry that has the HOMO-LUMO energy gap of 1.61 eV, in contrast to 1.71 eV  obtained by using the ab initio local-density approximation.

In this paper, persistent homology is, for the first time, employed to  {quantitatively} predict the stability of the fullerene molecules. The ground-state structures of a few  small fullerene molecules are first studied using a distance based filtration process. Essentially, we associate each carbon atom of a fullerene with an ever-increasing radius and thus define a Vietoris-Rips complex. The calculated Betti numbers (i.e., ranks of homology groups), including $\beta_0$, $\beta_1$ and $\beta_2$, are provided in the barcode representation. To further exploit the persistent homology, we carefully discriminate between the local short-lived and global long-lived bars in the barcodes.  We define an average accumulated bar length as the negative arithmetic mean of $\beta_2$ bars. As the local $\beta_2$ bars represent the number of cavities  of the structure, when $\beta_2$ becomes larger, interconnectedness (and thus stability) tends to increase, and relative energy tends to drop. Therefore, the average accumulated bar length indicates the level  of a relative energy. We validate this hypothesis with a series of ground-state structures of small fullerenes. It is found that our average accumulated bar length can capture the energy behavior remarkably well, including an anomaly in fullerene C$_{60}$ energy. Additionally, we explore the relative stability of fullerene isomers.  The persistence of the Betti numbers is calculated and analyzed. Our results are validated with the  total curvature energies of two fullerene families. It is observed that the total curvature energies of fullerene isomers can be  well represented with their  lengths of the long-lived Betti-2 bars, which indicates the sphericity of fullerene isomers. For fullerenes C$_{40}$ and C$_{44}$, correlation coefficients up to 0.956 and 0.948 are attained in the distance based filtration. Based on the flexibility-rigidity index (FRI) \cite{KLXia:2013d,KLXia:2013f,KLXia:2014b},  {a correlation matrix based filtration process is proposed to validate our findings}.

The rest of this paper is organized as follows. In Section \ref{Sec:theory}, we discuss the basic persistent homology concepts, including simplices and simplicial complexes, chains, homology, and filtration. Section \ref{Sec:algorithm} is devoted to the  description of algorithms. The alpha complex and Vietoris-Rips complex are discussed in some detail, including filtration construction, metric space design, and persistence evaluation. In Section \ref{sec:application}, persistent homology is employed in the analysis of fullerene structure and stability. After a brief discussion of fullerene structural properties, we elaborate on their barcode representation. The average accumulated bar length is introduced and applied to the energy estimate of the small fullerene series.  By validating with total curvature energies, our persistent homology based quantitative predictions are shown to be accurate.  Fullerene isomer  stability is also analyzed by using the new  correlation matrix based filtration.  This paper ends with a conclusion.

\section{Rudiments of Persistent Homology} \label{Sec:theory}

As representations of topological features, the homology groups are abstract abelian groups, which may not be robust or  {able to provide} continuous measurements. Thus, practical treatments of noisy data require the theory of persistent homology,
which provides  {continuous measurements} for the persistence of topological structures, allowing both quantitative comparison and  {noise removal} in topological analyses.
 {The concept was introduced by Frosini and Landi~\cite{Frosini:1999} and  Robins~\cite{Robins:1999},  and in the general form by Zomorodian and Carlsson~\cite{Zomorodian:2005}. Computationally, the first efficient algorithm for Z/2 coefficient situation was proposed by Edelsbrunner et al.~\cite{Edelsbrunner:2002} in 2002.}

\subsection{Simplex and Simplicial Complex}
For discrete surfaces, i.e., meshes, the commonly used homology is  {called simplicial homology}. To describe this notion, we first present a formal description of the meshes, the common discrete representation of surfaces and volumes. Essentially, meshing is a process in which  a  geometric shape is decomposed into elementary pieces called cells, the simplest of which are called \emph{simplices}.

\paragraph{Simplex}
 {\emph{Simplices}} are the simplest polytopes in a given dimension, as described below.  {Let $v_0,v_1,..v_p$ be $p\!+\!1$ affinely independent points in a linear space. A $p$-simplex $\sigma_p$ is the convex hull of those $p\!+\!1$ vertices}, denoted as  {$\sigma_p={\rm convex}<v_0,v_1,...,v_p>$} or shorten as  {$\sigma_p=<v_0,v_1,...,v_p>$}. 
 {A formal definition can be given as,}
\begin{eqnarray}
\sigma_p=\left\{v \mid v=\sum\limits_{i=0}^p\lambda_iv_i, \sum\limits_{i=0}^p\lambda_i=1, 0\leq\lambda_i\leq1, \forall i \right\}.
\end{eqnarray}

\begin{figure}
\vspace{-.1in}
	\centering
	\includegraphics[keepaspectratio,width=3.0in]{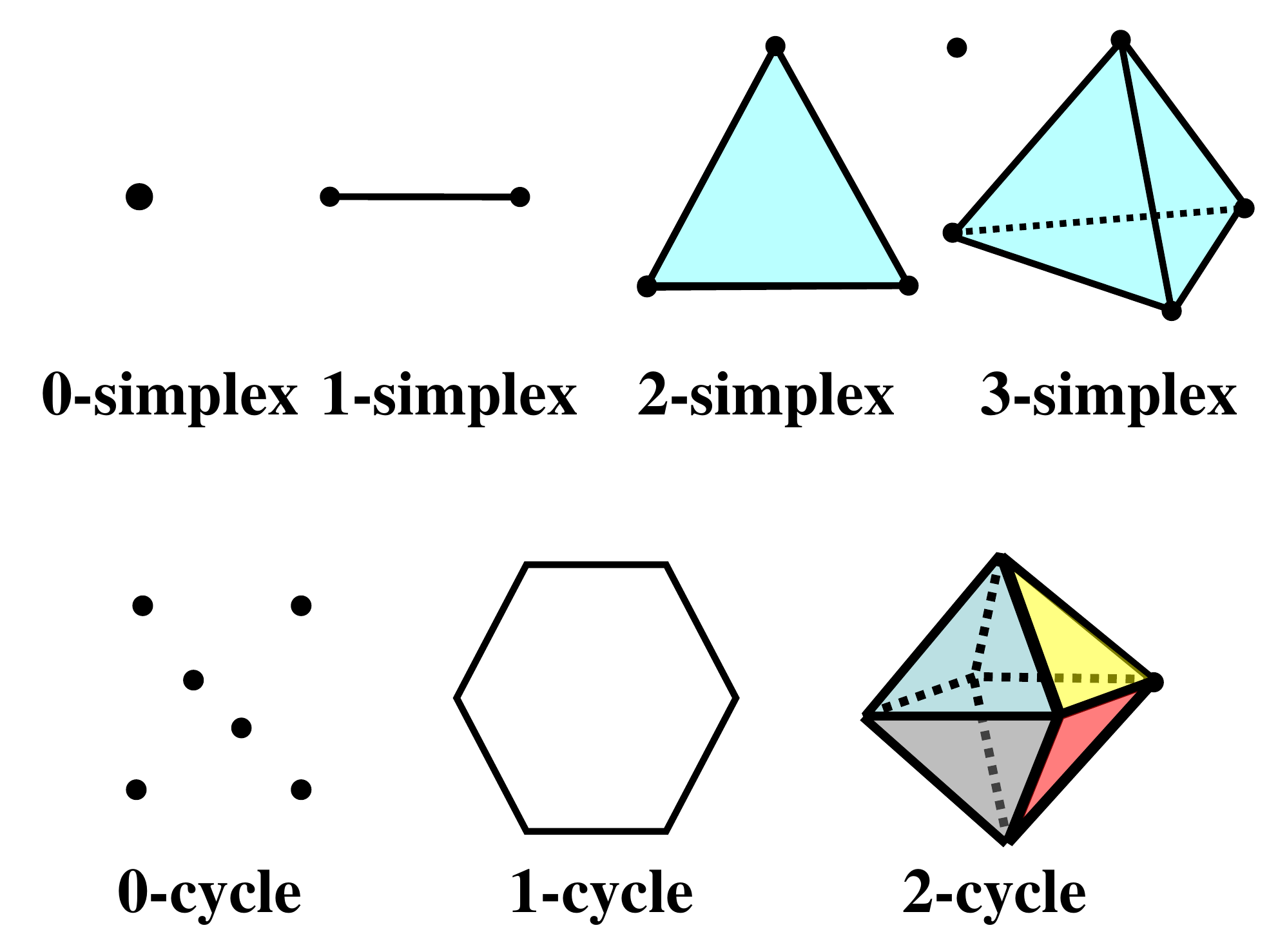}
	\caption{Illustration of 0-simplex, 1-simplex, and 2-simplex in the first row. The second row is simple 0-cycle, 1-cycle and 2-cycle.}
\label{fig:simplex}
	\vspace{-.1in}
\end{figure}
 {The most commonly used simplices in $\mathbb{R}^3$ are 0-simplex (vertex), 1-simplex (edge), 2-simplex (triangle) and 3-simplex (tetrahedron) as illustrated in Fig. \ref{fig:simplex}.}

An $m$-face of $\sigma_p$ is the $m$-dimensional subset of  {$m\!+\!1$} vertices, where $0\leq m\leq p$. For example, an edge has two vertices as its 0-faces and one edge as its 1-face. Since the number of subsets of a set with $p\!+\!1$ vertices is $2^{p\!+\!1}$, there are a  total of $2^{p\!+\!1}-1$ faces in $\sigma_p$.   {All the faces are proper except for $\sigma_p$ itself.} 
Note that polytope shapes can be decomposed into cells other than simplices, such as hexahedron and pyramid.  {However, as non-simplicial cells can be further decomposed, we can, without loss of generality,} restrict our discussion to shapes decomposed to simplices as we describe next.

\paragraph{Simplicial Complex}
With simplices as the basic building blocks, we define a \emph{simplicial complex} $K$ as a finite collection of simplices that meet the following two requirements,

\begin{itemize}
\item Containment: Any face of a simplex from $K$ also belongs to $K$.
\item Proper intersection: The intersection of any two simplices $\sigma_i$ and $\sigma_j$ from $K$ is either empty or a face of both $\sigma_i$ and $\sigma_j$.
\end{itemize}

Two $p$-simplices $\sigma^{i}$ and $\sigma^{j}$ are \emph{adjacent} to each other if they share a common face. The boundary of $\sigma_p$, denoted as $\partial{\sigma_p}$, is the union (which can be written as a formal sum) of its $(p\!-\!1)$-faces. Its interior is defined as the set containing  {all non-boundary points}, denoted as $\sigma-\partial{\sigma_p}$. We define a boundary operator for each $p$-simplex spanned by vertices $v_0$ through $v_p$ as
\begin{eqnarray}
\delta p<v_0,...,v_p>=\sum_{i=0}^p<v_0,...,\hat{v_i},...,v_p>,
\end{eqnarray}
where $\hat{v_i}$ indicates that $v_i$ is omitted  {and $Z/2$ coefficient set is employed}. It is the boundary operator that creates the nested topological structures and the
homomorphism among them as described in the next section.

If the vertex positions in the ambient linear space can be ignored or do not exist, the containment relation among the simplices (as finite point sets) defines an \emph{abstract simplicial complex}.

\subsection{Homology}\label{Homology}
A powerful tool in topological analysis is  {homology},  {which represents certain structures in the meshes by algebraic groups to describe their topology}. For regular objects in 3D space, essential topological features are connected components, tunnels and handles, and cavities, which are exactly described by the 0th, 1st, and 2nd  {homology groups}, respectively.

\paragraph{Chains}
The shapes to be mapped to  homology groups are constructed from \emph{chains} defined below.
Given a simplicial complex (e.g., a tetrahedral mesh) $K$, which, roughly
speaking,  {is a concatenation of $p$-simplices}
, we define a $p$-chain $c=\sum_i a_i \sigma_i$ as a formal linear
combination of all $p$-simplices in $K$,  {where $a_i\in Z/2$ is}
$0$ or $1$ and $\sigma_i$ is a $p$-simplex.  {Under such a definition, a
0-chain is a set of vertices, a 1-chain is a set of line segments which link vertices, a 2-chain is a set of triangles which are enclosed by line segments, and a 2-chain is a set of tetrahedrons which are enclosed by triangle surfaces.}

We extend the boundary operator $\partial_p$ for each $p$-simplex to a linear operator applied to chains, i.e., the extended operator meet following two conditions for linearity,
\begin{eqnarray}
\begin{aligned}
\partial_p(\lambda c)&=\lambda\partial_p(c),\\
\partial_p(c_i+c_j)&=\partial_p(c_i)+\partial_p(c_j),
\end{aligned}
\end{eqnarray}
where $c_i$ and $c_j$ are both chains and $\lambda$ is a constant, and all arithmetic is for modulo-2 integers, in which $1+1=0$.

An important property of the boundary operator is  {that the following composite operation is the zero map},
\begin{eqnarray}\label{BoundaryOperator}
\partial_p\circ\partial_{p+1}=0,
\end{eqnarray}
which immediately follows from the definition.
 {Take the 2-chain} $c=f_1+f_2$ as an example, which represents a membrane formed by two triangles, $f_1=<v_1, v_2, v_3>$ and  $f_2=<v_3, v_2, v_4>$. The boundary of $c$ is a 1-chain, which turns out to be a loop,
\begin{eqnarray}
\begin{aligned}
\partial_2(c)&=<v_1,v_2>+ <v_2,v_3> + <v_3, v_1> + <v_3, v_2> + <v_2, v_4> + <v_4, v_3>\\
&=<v_1,v_2>+ <v_3, v_1> + <v_2, v_4> + <v_4, v_3>.
\end{aligned}
\end{eqnarray}
The boundary of this loop is thus
\begin{eqnarray}
\begin{aligned}
\partial_1\circ\partial_2(c)&= \partial_1 (<v_1,v_2>+ <v_3, v_1> + <v_2, v_4> + <v_4, v_3>)\\
&= v_1 + v_2 +v_2 + v_4 + v_4 + v_3 + v_3 + v_1 = 0.
\end{aligned}
\end{eqnarray}

\paragraph{Simplicial homology}
Simplicial homology is built on the \emph{chain complex}  {associated to} the simplicial complex. A chain complex is a sequence of abelian groups $(C_1, C_2, \dots, C_n)$ connected by the homomorphism (linear operators) $\partial_p$, such that $\partial_p \circ \partial_{p+1} = 0$ as in Eq.(\ref{BoundaryOperator}).
\begin{eqnarray}
\cdots\xlongrightarrow{\partial_{p+1}}C_{p}\xlongrightarrow{\partial_{p}}C_{p-1}
\xlongrightarrow{\partial_{p-1}}\cdots\xlongrightarrow{\partial_{2}}C_{1}\xlongrightarrow{\partial_{1}}C_{0}
\xlongrightarrow{\partial_{0}}\emptyset.
\end{eqnarray}
The chain complex in the definition of simplicial homology is formed by C$_p$, the space of all $p$-chains, and $\partial_p$, the boundary operator on $p$-chains. Since $\partial_p\circ \partial_{p+1} = 0$, the kernel of the boundary operator $p$-chains is a subset of the image of the boundary operator of $p\!+\!1$-chains.
The $p$-chains in the kernel of the boundary homomorphisms $\partial_p$ are called \emph{$p$-cycles} ($p$-chains without boundary) and the $p$-chains in the image of the boundary homomorphisms $\partial_{p+1}$ are called $p$-boundaries.
The $p$-cycles form an abelian group (with group  {operation} being the addition of chains) called cycle group, denoted as $Z_p={\rm Ker}\ \partial_p$. The $p$-boundaries form another abelian group called boundary group, denoted as $B_p={\rm Im}\ \partial_{p+1}$.

\begin{figure}
\vspace{-.1in}
	\centering
	\includegraphics[keepaspectratio,width=5.0in]{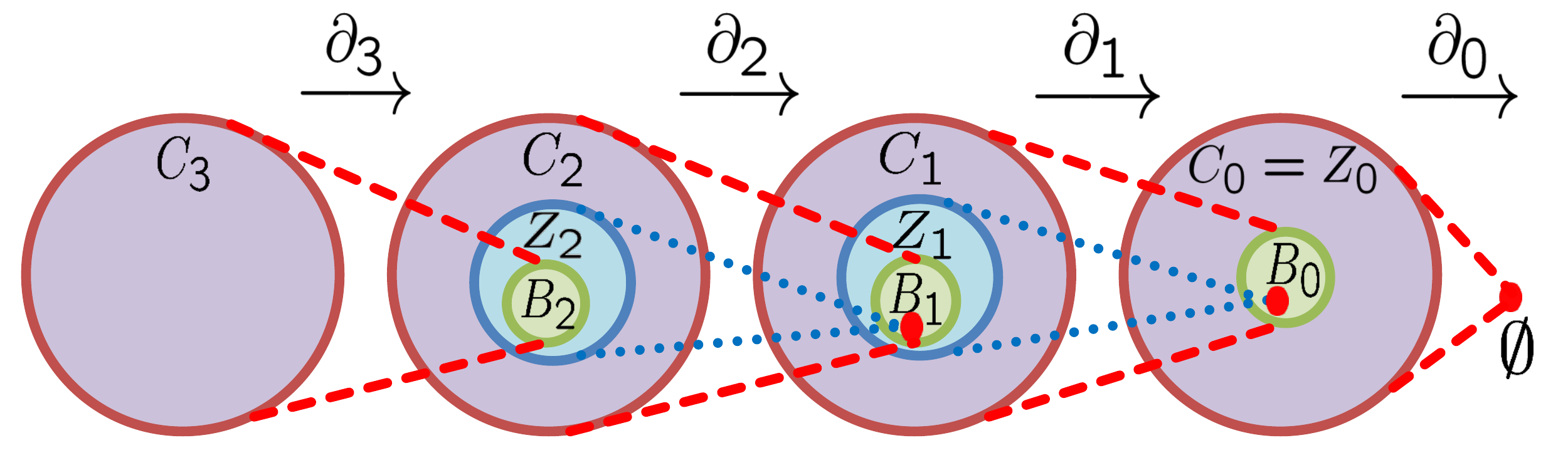}
	\caption{Illustration of boundary operators, and  chain, cycle and boundary groups  in $\mathbb{R}^3$. Red dots stand for empty sets.}
\label{fig:Homology}
	\vspace{-.1in}
\end{figure}

Thus, $p$-boundaries are also $p$-cycles as shown in Fig.    \ref{fig:Homology}. As $p$-boundaries form a subgroup of the cycles group, the quotient group can be constructed through cosets of $p$-cycles, i.e., by equivalence classes of cycles. The $p$-th homology, denoted as $H_p$, is defined as a quotient group,
\begin{eqnarray}
\begin{aligned}
H_p&={\rm Ker}\ \partial_p/{\rm Im}\ \partial_{p+1}\\
&=Z_p/B_p,
\end{aligned}
\end{eqnarray}
where ${\rm Ker}\ \partial_p$ is the collection of $p$-chains with empty boundary and ${\rm Im}\ \partial_{p+1}$
is the collection of $p$-chains that are boundaries of $p+1$-chains.

 {Noticing that all groups with $p > 3$ cannot be generated from meshes in $\mathbb{R}^3$}, we only need chains, cycles and boundaries of dimension $p$ with $0\leq p \leq 3$. See Fig.    \ref{fig:Homology} for an illustration.

 {We illustrate simplexes and cycles including 0-cycle, 1-cycle, and 2-cycle in Fig. \ref{fig:simplex}. Basically,} an element in the $p$-th homology group is an equivalence class of $p$-cycles. One of these cycles $c$ can represent any other $p$-cycle that can be ``deformed'' through the mesh to $c$, because any other $p$-cycle in the same equivalence class differ with $c$ by a $p$-boundary $b=\partial (\sigma_1+\sigma_2+\dots)$, where each $\sigma_i$ is a $p\!+\!1$-simplex. Adding the boundary of $\sigma_i$ has the effect of deforming $c$ to $c+\partial\sigma_i$ by sweeping through $\sigma_i$. For instance, a $0$-cycle $v_i$ is equivalent to $v_j$ if there is a path $<v_i, v_{k1}> + <v_{k1},v_{k2}> + \dots + <v_{kn}, v_j>$. Thus each generator  of  $0$th-homology, (like a basis vector in a basis of the linear space of $0$th-homology) represents one connected component. Similarly, $1$-cycles are loops, and
$1$st-homology generators represent independent nontrivial loops, i.e., separate tunnels;
$2$-homology generators are independent membranes, each
enclosing one cavity of the 3D object.

Define $\beta_p={\rm rank}(H_p)$ to be the $p$-th Betti number. For a simplicial complex in 3D, $\beta_0$ is the number of connected components;
$\beta_1$ is the number of tunnels; and
$\beta_2$ is the number of cavities.
As $H_p$ is the quotient group between $Z_p$ and $B_p$, we can also compute Betti numbers through,
\begin{equation}
{\rm rank}(H_p) = {\rm rank}(Z_p) - {\rm rank}(B_p).
\end{equation}
Note, however, $H_p$ is usually of much lower rank than either $Z_p$ or $B_p$.

\subsection{Persistent Homology}

Homology generators identify the tunnels, cavities, etc., in the shape, but as topological invariants, they omit the metric measurements by definition. However, in practice, one often needs to compare the sizes of tunnels, for instance, to find the narrowest tunnel, or to  {remove} tiny tunnels as topological noises. Persistent homology is a method of reintroducing metric measurements to the topological structures~\cite{Edelsbrunner:2002,Zomorodian:2005}.

The measurement is introduced as an index $i$ to a sequence of nested topological spaces $\{\mathbb{X}_i\}$. Such a sequence is called a \emph{filtration},
\begin{equation}
\emptyset = \mathbb{X}_0\subseteq \mathbb{X}_1\subseteq \mathbb{X}_2\subseteq \cdots \subseteq \mathbb{X}_m = \mathbb{X}.
\end{equation}
Since each inclusion induces a mapping of chains, it induces a linear map for homology,
\begin{equation}
\emptyset= H(\mathbb{X}_0)\rightarrow H(\mathbb{X}_1)\rightarrow H(\mathbb{X}_2)\rightarrow \cdots \rightarrow H(\mathbb{X}_m) = H(\mathbb{X}).
\end{equation}

\begin{figure}
\begin{center}
\vspace{-0.in}
\includegraphics[keepaspectratio,width=5.0in]{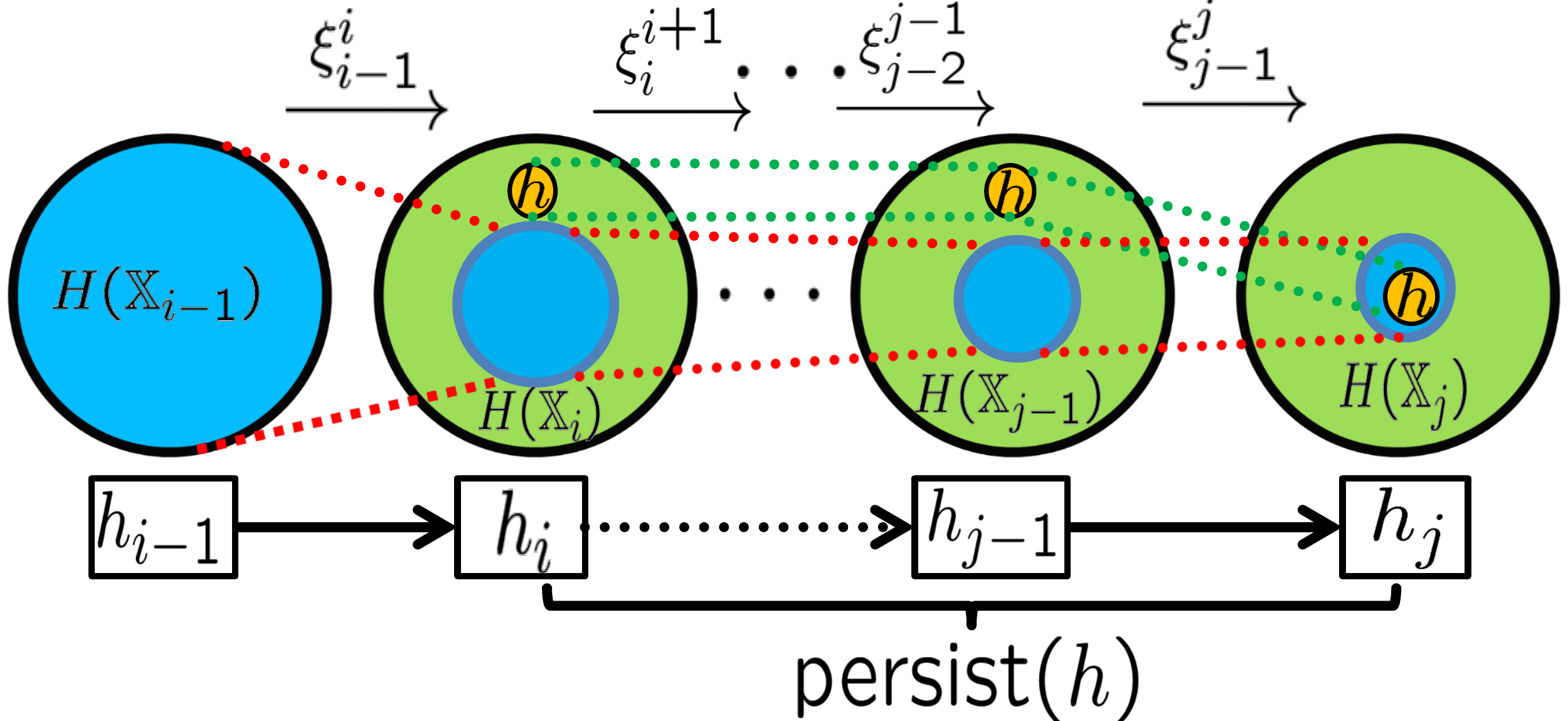}
	\vspace{-0.in}
\end{center}
\caption{Illustration of the birth and death of a homology generator $c$}
\label{fig:Filtration}
\end{figure}
The above sequence describes the evolution of the homology generators. We follow the exposition in Ref. \cite{Munch:2013}
and  {define by} a composition mapping from $H(\mathbb{X}_i)$ to $H(\mathbb{X}_j)$ as
$\xi_i^j: H(\mathbb{X}_i)\rightarrow H(\mathbb{X}_j)$.
A new homology class $c$ is created (born) in $\mathbb{X}_i$ if it is not in the image of $\xi_{i-1}^i$.
It dies in $\mathbb{X}_j$ if it becomes trivial or is merged to an ``older'' (born before $i$) homology class, i.e., its image in $H(\mathbb{X}_j)$ is in the image of $\xi_{i-1}^j$,
unlike its image under $\xi_{i}^{j-1}$.

As shown in Fig.   ~\ref{fig:Filtration}, if we associate with each space $\mathbb{X}_i$ a value $h_i$ denoting ``time'' or ``length'', we can define the duration, or the persistence length of the each homology generator $c$ as
\begin{equation}
{\rm persist}(c) = h_j-h_i.
\end{equation}
This measurement $h_i$ is usually readily available when analyzing the topological feature changes. For instance, when the filtration arises from the level sets of a height function.

\section{Algorithms for persistent homology}\label{Sec:algorithm}
\begin{figure}
\begin{center}
\begin{tabular}{c}
\includegraphics[width=1\textwidth]{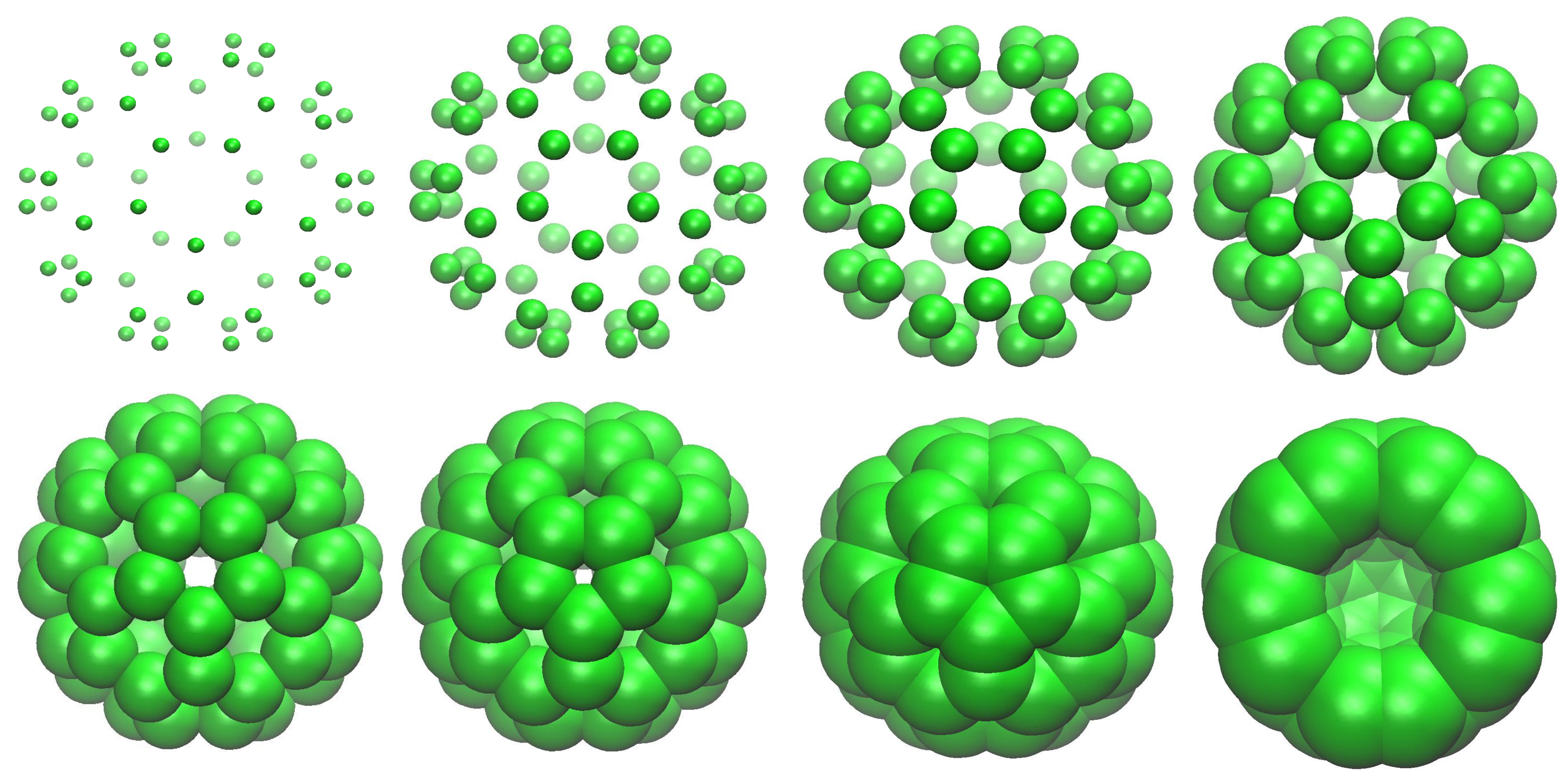}
\end{tabular}
\end{center}
\caption{Illustration of filtrations built  on fullerene C$_{60}$. Each point or atom in the point cloud data (i.e., coordinates) of the C$_{60}$ is associated with a common radius $r$ which increases gradually. As the value of $r$ increases, the solid balls centered at given coordinates grow.  These balls eventually overlap with their neighbors at certain $r$ values. Simplices indicating such neighborhood information can be defined through abstract $r$-dependent simplicial complexes, e.g., alpha complexes and Rips complexes.  {Note that in the last chart, we have removed some atoms to reveal the central void.}}
\label{fig:carbon60}
\end{figure}
In computational topology, intrinsic features of point cloud data, i.e., a point set $S\subset \mathbb{R}^n$ without additional structure, are common subjects of investigation. For such data, a standard way to construct the filtration is to grow a solid ball centered at each point with an ever-increasing radius. If the differences between points can generally be ignored, as is the case for fullerenes, a common radius $r$ can be used for all points. In this setting, the radius $r$ is used as the parameter for the family of spaces in the filtration. As the value of $r$ increases, the solid balls will grow and simplices can be defined through the overlaps among the set of balls.
In Figure \ref{fig:carbon60}, fullerene C$_{60}$ is used to demonstrate this process. There are various ways of constructing abstract simplicial complexes from the intersection patterns of the set of expanding balls, such as \v{C}ech complex, Vietoris-Rips complex and alpha complex. The corresponding topological invariants, e.g., the Betti numbers, are in general different due to different definitions of simplicial complexes. In this section, we discuss computational algorithms for the design of filtrations, the construction of abstract simplicial complexes, and the calculation of Betti numbers.

\paragraph{Alpha complex}

One possible filtration that can be derived from the unions of the balls with a given radius around the data points (as shown in Figure \ref{fig:carbon60}) is the family of $d$-dependent \v{C}ech complexes, each of them is defined to be a simplicial complex, whose $k$-simplices are determined by $(k+1)$-tuples of points, such that the corresponding $d/2$-balls have a non-empty intersection. However, it may contain many simplices for a large $d$. A variant called  {the alpha complex} can be defined by replacing the $d/2$-ball in the above definition by the intersection of the  {$d/2$-ball with the Voronoi cells for these data points}. {In both cases, they are homotopic to the simple unions of balls, and thus produce the same persistent homology.  {Interested readers are referred to the nerve theorem for details \cite{Zomorodian:2009book}.}

\paragraph{Vietoris-Rips complex}
The Vietoris-Rips complex, which is also known as Vietoris complex or Rips complex, is another type of abstract simplicial complex derived from the union of balls. In this case, for a $k$-simplex to be included, instead of requiring that the $(k+1)$ $d/2$-balls to have a common intersection, one only needs them to intersect pairwise. The \v{C}ech complex is a subcomplex of the Rips complex for any given $d$, however, the latter is much easier to compute and is also a subcomplex of the former at the filtration parameter of $\sqrt{2}d$.

\paragraph{Euclidean-distance based filtration}
It is straightforward to use the metric defined by the Euclidean space in which the data points are embedded. The pairwise distance can be stored in a symmetric distance matrix $\left(d_{ij}\right)$, with each entry $d_{ij}$ denoting the distance between point $i$ and point $j$. Each diagonal term of the matrix is the distance from a certain point to itself, and thus is always 0. The family of Rips complexes is parameterized by $d$, a threshold on the distance. For a certain value of $d$, the Vietoris-Rips complex can be calculated. In 3D, more specifically, for a pair of points whose  distance is below the threshold $d$, they form  {a 1-simplex} in the Rips complex; for a triplet of points,  {if the distance between every pair is smaller than $d$}, the 2-simplex formed by the triplet is in the Rips complex; whether a 3-simplex is in the Rips complex can be similarly determined. The Euclidean-distance based Vietoris-Rips complexes are widely used in persistent homology due to their simplicity and efficiency.

\paragraph{Correlation matrix based filtration}
\begin{figure}
\begin{center}
\begin{tabular}{c}
\includegraphics[width=0.8\textwidth]{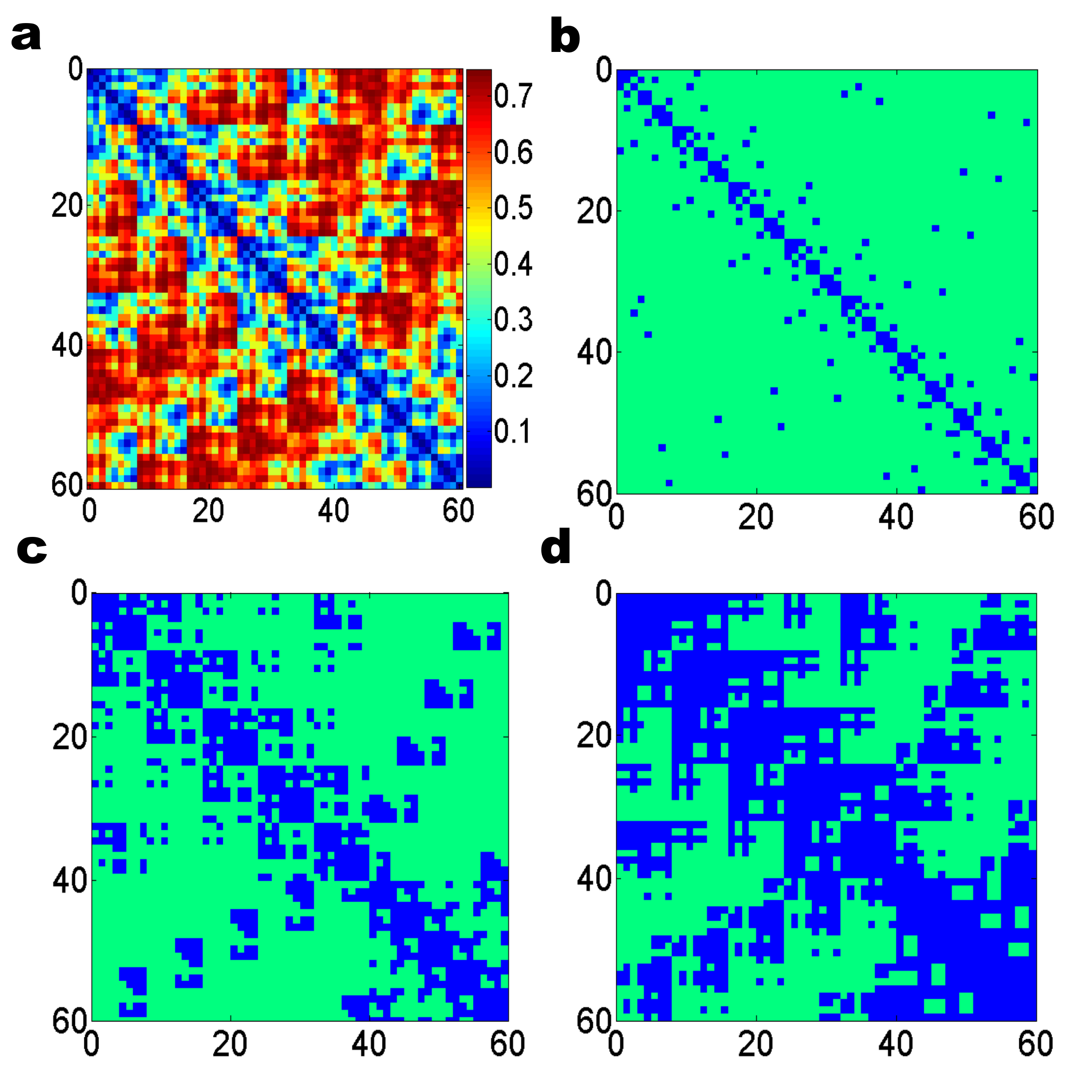}
\end{tabular}
\end{center}
\caption{Correlation matrix based filtration of fullerene C$_{60}$ (labels on both axes are atomic numbers). A correlation matrix is constructed from the FRI theory. As the filtration parameter increases, the Rips complex based on this matrix expands accordingly. ({\bf a}) The correlation based matrix for fullerene C$_{60}$; ({\bf b}), ({\bf c}) and ({\bf d}) demonstrate the connectivity between atoms at the filtration threshold $d=0.1$\AA, $0.3$\AA, and $0.5$\AA, respectively. The blue color entries represent the pairs already forming simplices. }
\label{fig:FiltrationMatrix}
\end{figure}
Another way to construct the metric space is through a certain correlation matrix, which can be built, e.g., from theoretical predictions and experimental observations. From a previous study on protein stability, flexibility-rigidity index (FRI) theory has been proven accurate and efficient\cite{KLXia:2013d}. The reason for its success is that the geometric information is harnessed properly through the special transformation to a correlation matrix. The key to this transformation is the geometric to topological mapping. Instead of direct geometric information of the embedding in the Euclidean space, a mapping through certain kernel functions is able to interpret spatial locations of atoms in a particular way that reveals the atom stability quantitatively.  We believe that this functional characterization is of  importance to the study of not only proteins, but also other molecules.

Here, we present a special correlation {matrix based} Vietoris complex on the FRI method. In order to define the metric used, we briefly review the concepts of the FRI theory. First, we introduce the geometry to topology mapping \cite{KLXia:2013d,KLXia:2013f,KLXia:2014b}. We denote the coordinates of atoms in the molecule we study as ${\bf r}_{1}, {\bf r}_{2}, \cdots, {\bf r}_{j}, \cdots, {\bf r}_{N}$, where ${\bf r}_{j}\in \mathbb{R}^{3}$ is the position vector of the $j$th atom.  The Euclidean distance between $i$th and $j$th atoms $r_{ij}$ can then be calculated. Based on these distances, topological connectivity matrix can be constructed with monotonically decreasing radial basis functions. A general form for a  connectivity matrix is,
\begin{eqnarray}\label{eq:couple_matrix0}
{C}_{ij} = w_{j} \Phi( r_{ij},\eta_{j}),
\end{eqnarray}
where $w_{j}$ is associated with atomic types,  parameter $\eta_{j}>0$ is the atom-type related characteristic distance, and $\Phi( r_{ij};\eta_{j}) $ is a radial basis correlation kernel. 

The choice of kernel is of significance to the FRI model. It has been shown that highly predictive results can be obtained by
the exponential type and Lorentz type of kernels \cite{KLXia:2013d,KLXia:2013f,KLXia:2014b}.   Exponential type of kernels is
\begin{eqnarray}\label{eq:ExpKernel}
\Phi(r,\eta) =    e^{-\left(r/\eta\right)^\kappa},   \quad \eta >0,  \kappa >0
\end{eqnarray}
and the Lorentz type of kernels is
\begin{eqnarray}\label{eq:PowerKernel}
 \Phi(r, \eta) =
 \frac{1}{1+ (r/\eta)^{\upsilon}}.
  \quad \eta >0,  \upsilon > 0
 \end{eqnarray}
The parameters $\kappa$ and $\upsilon$ are adjustable. 

We  define the atomic rigidity index  $\mu_i$ for $i$th atom as
\begin{eqnarray}\label{eq:rigidity1}
 \mu_i = \sum_{j=1}^N w_{j} \Phi( r_{ij} ,\eta_{j} ), \quad \forall i =1,2,\cdots,N.
\end{eqnarray}
A related atomic flexibility index can be defined as the inverse of the atomic rigidity index.
\begin{eqnarray}\label{eq:flexibility1}
f_i= \frac{1}{\mu_i}, \quad \forall i =1,2,\cdots,N.
\end{eqnarray}

The FRI theory has been intensively validated by comparing with the experimental data, especially the Debye-Waller factor (commonly known as the B-factor) \cite{KLXia:2013d}.  While simple to evaluate, their applications in B-factor prediction yield decent results. The predicted results are proved to be highly accurate while the procedure remains efficient. FRI is also used to analyze the protein folding behavior \cite{KLXia:2014b}.

To construct an FRI-based metric space, we need to design a special distance matrix, in which the functional correlation is measured. If we directly employ the correlation matrix in Eq. (\ref{eq:couple_matrix0}) for the filtration, atoms with less functional relation  {form} more simplices, resulting in a counter-intuitive persistent homology. However, this problem can be easily remedied by defining a new correlation matrix as $M_{ij}=1-{C}_{ij}$, i.e.,
\begin{eqnarray}\label{eq:FiltrationMatrix}
{M}_{ij} = 1-w_{j} \Phi( r_{ij}, \eta_{j}).
\end{eqnarray}
Thus a kernel function induces a metric space under this definition. Figure \ref{fig:FiltrationMatrix}({\bf a}) demonstrates such a metric space based filtration  of  fullerene C$_{60}$, in which we assume $w_{j}=1$ since only one type of atom exists in this system. The generalized exponential kernel in Eq. (\ref{eq:ExpKernel}) is used with parameters $\kappa=2.0$ and $\eta=6.0$\AA.

With the correlation matrix based filtration, the corresponding Vietoris-Rips complexes can be straightforwardly constructed. Specifically, given a certain filtration parameter $h_0$, if the matrix entry $M_{ij}\leq h_0$, an edge formed between $i$th and $j$th atoms, and a simplex is formed if all of its edges are present. The complexes are built incrementally as the filtration parameter grows. Figures \ref{fig:FiltrationMatrix}({\bf b}), ({\bf c}) and ({\bf d}) illustrate this process with three filtration threshold values $h=0.1$\AA, $0.3$\AA~ and $0.5$\AA, respectively. We use the blue color to indicate formed edges. It can be seen that simplicial complexes keep growing with  the increase of filtration parameter $h$. The diagonal terms are always equal to zero, which means that  $N$ atom centers (0-simplices) form the first complex in the filtration.


\section{Application to fullerene structure  analysis and stability prediction}\label{sec:application}

In this section, the theory and algorithms of persistent homology are employed to study the structure and stability   of fullerene molecules. The ground-state structural data of fullerene molecules used in our tests are downloaded from the  \href{http://www.ccl.net/cca/data/fullerenes}{CCL webpage} and  fullerene  isomer data and corresponding total curvature energies \cite{Guan:2014} are adopted from David Tomanek's  \href{http://www.nanotube.msu.edu/fullerene}{carbon fullerene webpage}. In these structural data, coordinates of fullerene carbon atoms are given. The collection of atom center locations of each molecule forms a point cloud in $\mathbb{R}^{3}$. With only one type of atom, the minor heterogeneity of atoms due to their chemical environments in these point clouds can  be ignored in general. We examined both distance based and correlation matrix  based metric spaces in our study. The  filtration based on the FRI theory is shown to predict the stability very well.


Before we discuss the more informative persistent homology of fullerenes, we discuss the basic structural properties simply based  on their Euler characteristics  {(vertex number minus edge number plus polygon number)}. The Euler characteristic, as a topological property, is  invariant under non-degenerate shape deformation. For a fullerene cage composed of only pentagons and hexagons, the exact numbers of these two types of polygons can be derived from the Euler characteristic. For instance, if we have $n_p$ pentagon and $n_h$ hexagons in a C$_N$ fullerene cage, the corresponding numbers of vertices, edges and faces are
$(5n_p+6n_h)/3$, $(5n_p+6n_h)/2$ and $n_p+n_h$, respectively, since each vertex is shared by three faces, and each edge is shared by two faces. As the fullerene cage is treated as a two dimensional surface, we have the Euler characteristic  $(5n_p+6n_h)/3-(5n_p+6n_h)/2+(n_p+n_h)=2$, according to Euler's polyhedron formula, since it is a topological sphere. Thus, we have $n_p=12$, which means a fullerene cage structure must have 12 pentagons and correspondingly $N/2-10$ hexagons. Therefore, for a C$_N$ fullerene cage, we have $N$ vertices, $3N/2$ edges and $N/2+2$ faces.

\subsection{Barcode representation of fullerene structures  {and nanotube}}\label{Sec:FullereneBarcodes}
\begin{figure}
\begin{center}
\begin{tabular}{cc}
\includegraphics[width=0.48\textwidth]{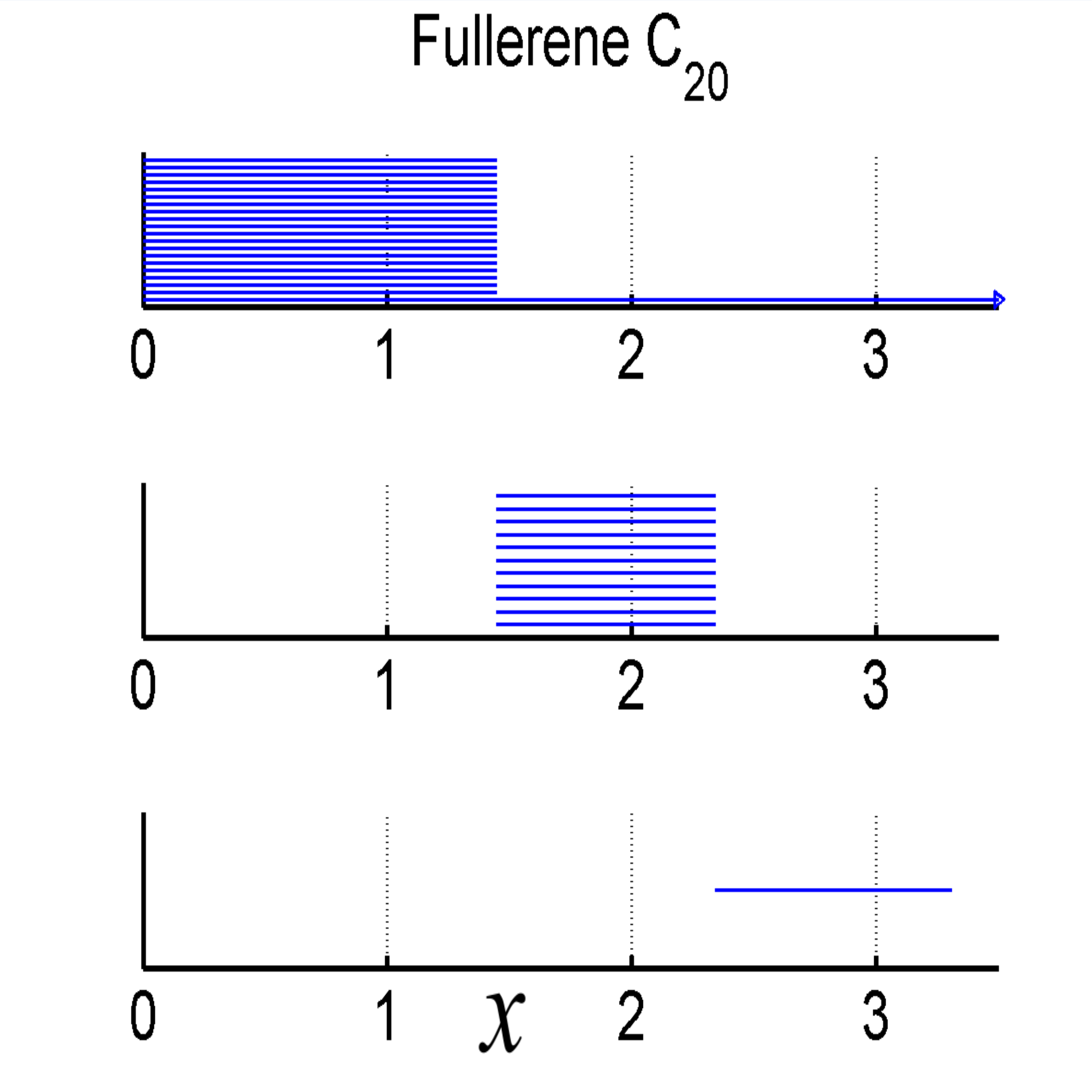}&
\includegraphics[width=0.48\textwidth]{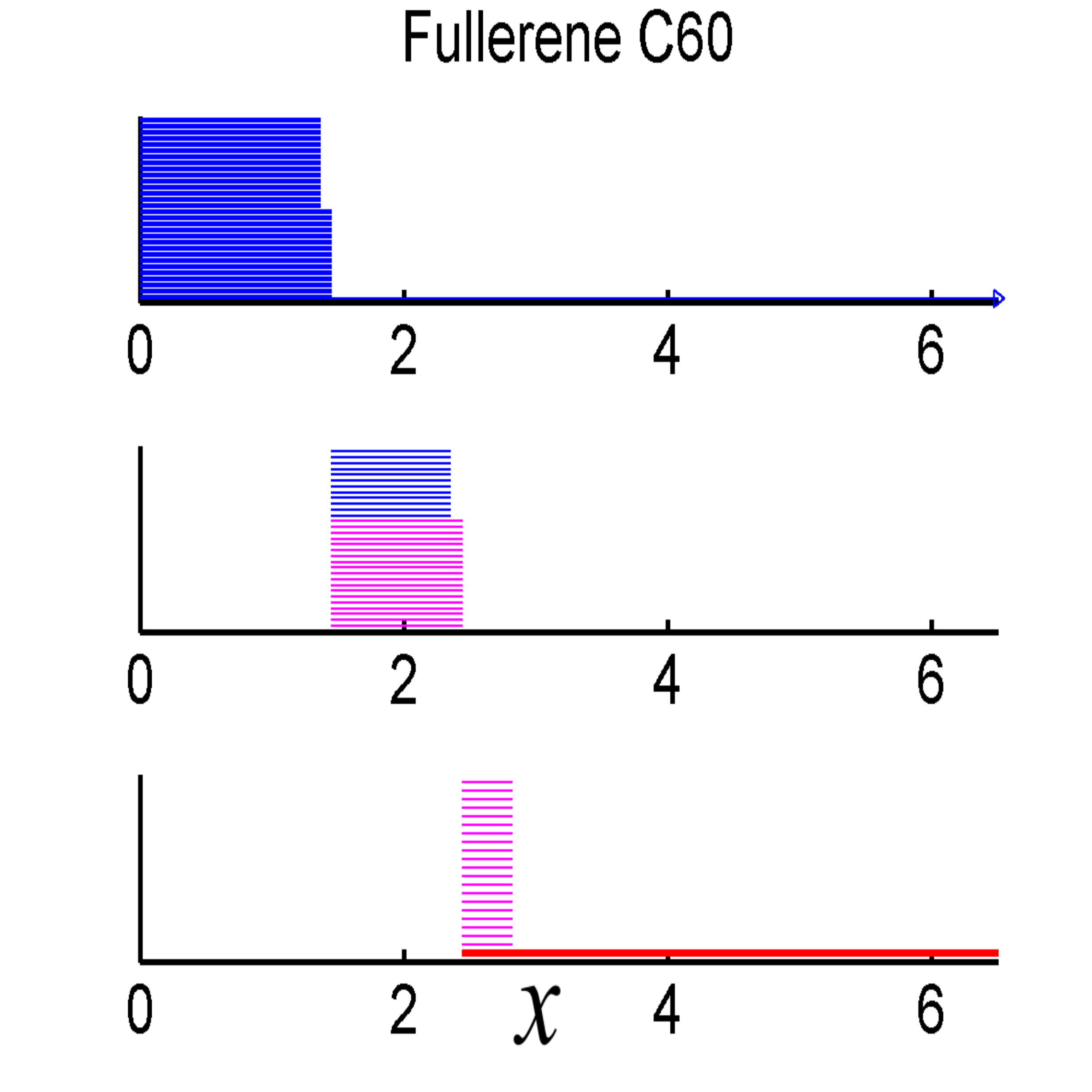}
\end{tabular}
\end{center}
\caption{Illustration of the barcodes for fullerene C$_{20}$(left chart) and C$_{60}$ (right chart) filtration on associated Rips complexes. Each chart contains three panels corresponding to the Betti number sequences $\beta_0,\beta_1$ and $\beta_2$, from top to bottom. 
}
\label{fig:C20C60Barcode}
\end{figure}

\paragraph{Barcodes for fullerene molecule}
{In Fig.    \ref{fig:C20C60Barcode}, we demonstrate the persistent homology analysis of fullerene C$_{20}$ and C$_{60}$ using the barcode representation generated by  {Javaplex \cite{javaPlex}.} }
 {The $x$-axis represents the filtration parameter $h$. If the distance between two vertices is below or equal to certain $h_0$, they will form an edge (1-simplex) at $h_0$. Stated differently, the simplical complex generated is equivalent to the raidus filtration with radius parameter $h/2$}. In the barcode, the persistence of a certain Betti number is represented by an interval (also known as bar), denoted as $L^{\beta_j}_{i}, j=0,1,2; i=1,2,\cdots$. Here $j\in \{0,1,2\}$ as we only consider the first three Betti numbers in this work. From top to bottom, the behaviors of $\beta_0$, $\beta_1$, and  $\beta_2$ are depicted in three individual panels. It is seen that as $h$ grows, isolated atoms initialized as points will gradually grow into solid spheres with an ever-increasing radius. This phenomenon  is represented by the bars in the $\beta_0$ panel. Once two spheres overlap with each other, one $\beta_0$ bar is terminated. Therefore, the bar length for the independent 0-th homology generator (connected component) $c^0_i$, denoted as $L^{\beta_0}_{i}={\rm persist}(c^0_i)$, indicates the bond length information of the molecule. As can be seen from Fig.    \ref{fig:C20C60Barcode}, for fullerene C$_{20}$, all $\beta_0$ bar lengths are around $1.45$\AA~ and the total number of components equals exactly to 20. On the other hand, fullerene C$_{60}$ has two different kinds of bars with lengths around $1.37$\AA~ and $1.45$\AA, respectively, indicating its two types of bond lengths.

More structure information is revealed as $\beta_1$ bars, which represent independent noncontractible $1$-cycles (loops), emerge. It is seen in the fullerene C$_{20}$ figure, that there are 11 equal-length $\beta_1$ bars persisting from $1.45$\AA~ to $2.34$\AA. As fullerene C$_{20}$ has 12  {pentagonal rings},  {the Euler characteristics for a 1D simplicial subcomplex ($1$-skeleton) can be evaluated from the Betti numbers},
\begin{equation}\label{Euler_1simplicial}
n_{\rm vertice}-n_{\rm edge}=\beta_0-\beta_1.
\end{equation}
Here $\beta_0$, $n_{\rm vertice}$, and $n_{\rm edge}$ are 1, 20, and 30, respectively. Therefore, it is easy to obtain that $\beta_1=11$ for  fullerene C$_{20}$,  as demonstrated in Fig.    \ref{fig:C20C60Barcode}. It should be noticed that all $\beta_1$ bars end at filtration value $h=2.34$\AA, when five balls in each pentagon with their ever-increasing radii begin to overlap to form a pentagon surface.

Even more structural information can be derived from fullerene C$_{60}$'s $\beta_1$ barcodes. First, there are $31$ bars for $\beta_1$. This is consistent with the Euler characteristics in Eq. (\ref{Euler_1simplicial}), as we have 12 pentagons and 20 hexagons.  Secondly, two kinds of bars correspond to the coexistence of  {pentagonal rings} and  {hexagonal rings}. They persist from $1.45$\AA~  to $2.35$\AA~ and from $1.45$\AA~ to $2.44$\AA~, respectively.

As the filtration progresses, $\beta_2$ bars (membranes enclosing cavities) tend to appear. In fullerene C$_{20}$, there is only one $\beta_2$ bar, which corresponds to the void structure in the center of the cage. For fullerene C$_{60}$, we have 20 $\beta_2$  {bars persisting} from  $2.44$\AA~ to $2.82$\AA,  {which corresponds to hexagonal cavities as indicated in the last chart of Fig .\ref{fig:simplex}.  Basically, as the filtration goes, each node in the hexagon ring joins its four nearest neighbors, and fills in the relevant 2-simplices, yielding a simplical complex whose geometric realization is exactly the octahedron.} There is another  $\beta_2$ bar due to the center void  {as indicated in the last chart of Fig.\ref{fig:C20C60Barcode}}, which persists until the complex forms a solid block. Note that two kinds of $\beta_2$ bars represent entirely different physical properties. The short-lived bars  are related to local behaviors and fine structure details, while the long-lived bar is associated with the global feature, namely, the large cavity.

\begin{figure}
\begin{center}
\begin{tabular}{c}
\includegraphics[width=0.7\textwidth]{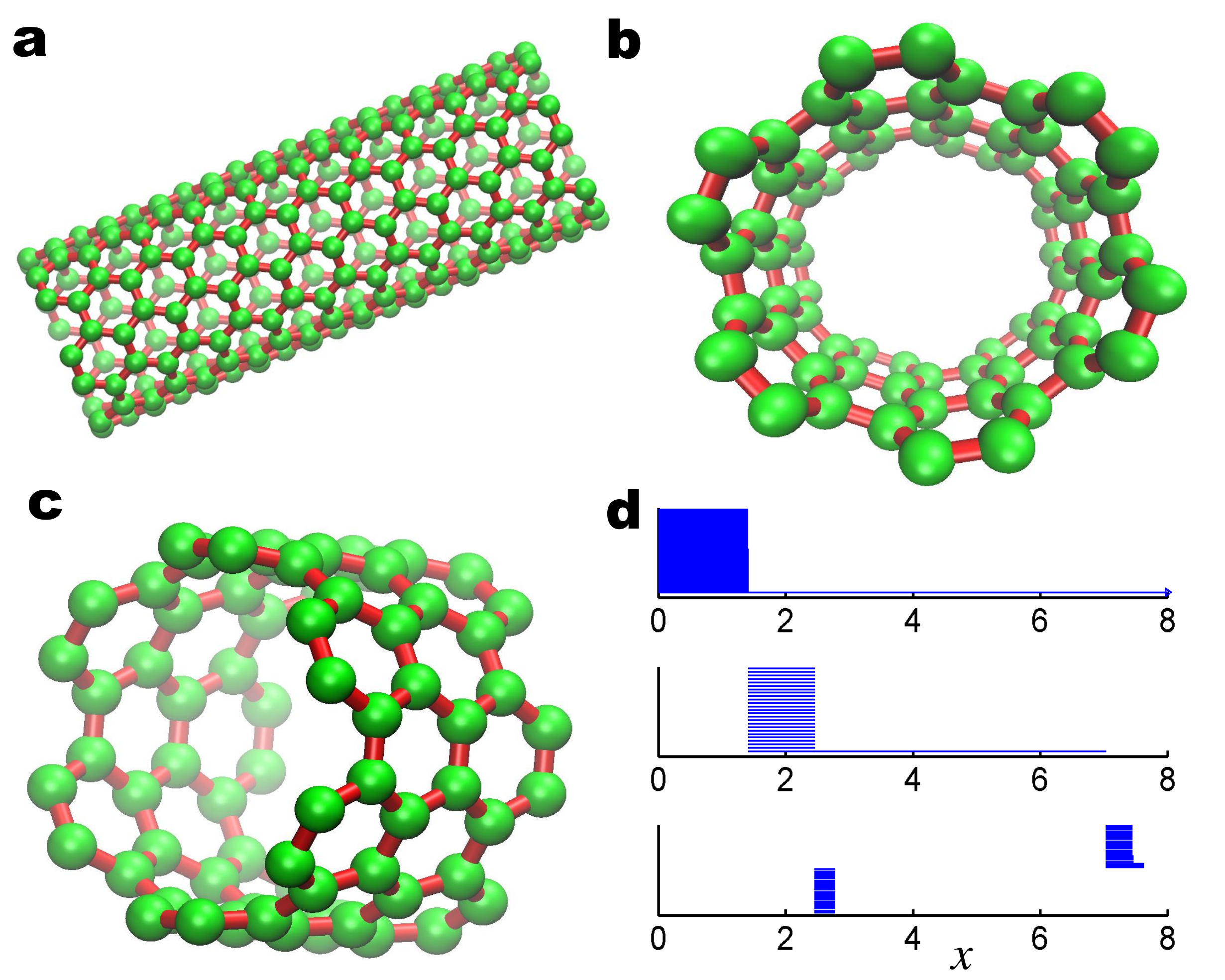}
\end{tabular}
\end{center}
\caption{Illustration of persistent homology analysis  for a nanotube. ({\bf a}) The generated nanotube structure with 10 unit layers. ({\bf b}) and ({\bf c}) A 3 unit layer segment extracted from the nanotube molecule in {\bf a}. ({\bf d}) Barcodes representation of the topology of the nanotube segment.}
\label{fig:nanotube}
\end{figure}

\paragraph{Barcodes for nanotube}
 {Another example of nanotube is demonstrated in Fig. \ref{fig:nanotube}. The nanotube structure is constructed using the software \href{https://www.ccs.uky.edu/~ernst/carbontubes/TubeApplet.html}{TubeApplet webpage}. We set tube indices to (6,6), the number of unit cell to 10, tube radius to 4.05888, and lattice constant to 2.454\AA. We extract a segment of 3 unit cells from the nanotube and employ the persistent homology analysis to generate it barcodes. Our results are demonstrated in Fig.   \ref{fig:nanotube}. Different from  {fullerene} molecules, the nanotube has a long $\beta_1$  bar representing the tube circle.}  {It should also be noticed that $\beta_2$ barcodes are concentrated in two different regions. The first region is when $x$ is around 2.5 to 2.7. The $\beta_2$ barcodes in this domain are generated by hexagonal rings on the nanotube. The other region appears when $x$ is slightly larger than 7.0. The corresponding $\beta_2$ barcodes are representation of the void formed between different layer of carbons. }

 {Unlike commonly used  topological methods\cite{Fowler:1995}}, persistent homology is able to provide a multiscale representation of the topological features. Usually, global behavior is of  {major concern}. Therefore, the importance of the topological features is typically measured by their persistence length.} In our analysis, we have observed that except for discretization errors, topological invariants of all scales can be equally important in revealing various structural features of the system of interest. In this work, we demonstrate that both  local and global topological invariants play important roles in quantitative physical modeling.

\subsection{Stability analysis of small fullerene molecules}

From the above analysis, it can be seen that detailed structural information has been incorporated into the corresponding barcodes. On the other hand, molecular structures determine  molecular functions \cite{KLXia:2013d,KLXia:2013f,KLXia:2014b}. Therefore, persistent homology can be used to predict molecular functions of fullerenes. To this end, we analyze the barcode information.  For each Betti number $\beta_j$, we define an accumulated bar length $A_j$ as the summation of barcode lengths,
\begin{equation}\label{AccumulationIndex}
A_j=\sum_{i=1} L^{j}_{i}, j=0,1,2,
\end{equation}
where  $L^j_{i}$ is the length of the $i$th bar in the $j$-th-homology barcode.  Sometimes, we may only sum over certain types of barcodes.
We define an  average accumulated bar length as $B_j=-\sum_{i=1} L^{j}_{i}/N$,  where $N$ is the total number of atoms in the molecule.

\begin{figure}
\begin{center}
\begin{tabular}{cc}
\includegraphics[width=0.45\textwidth]{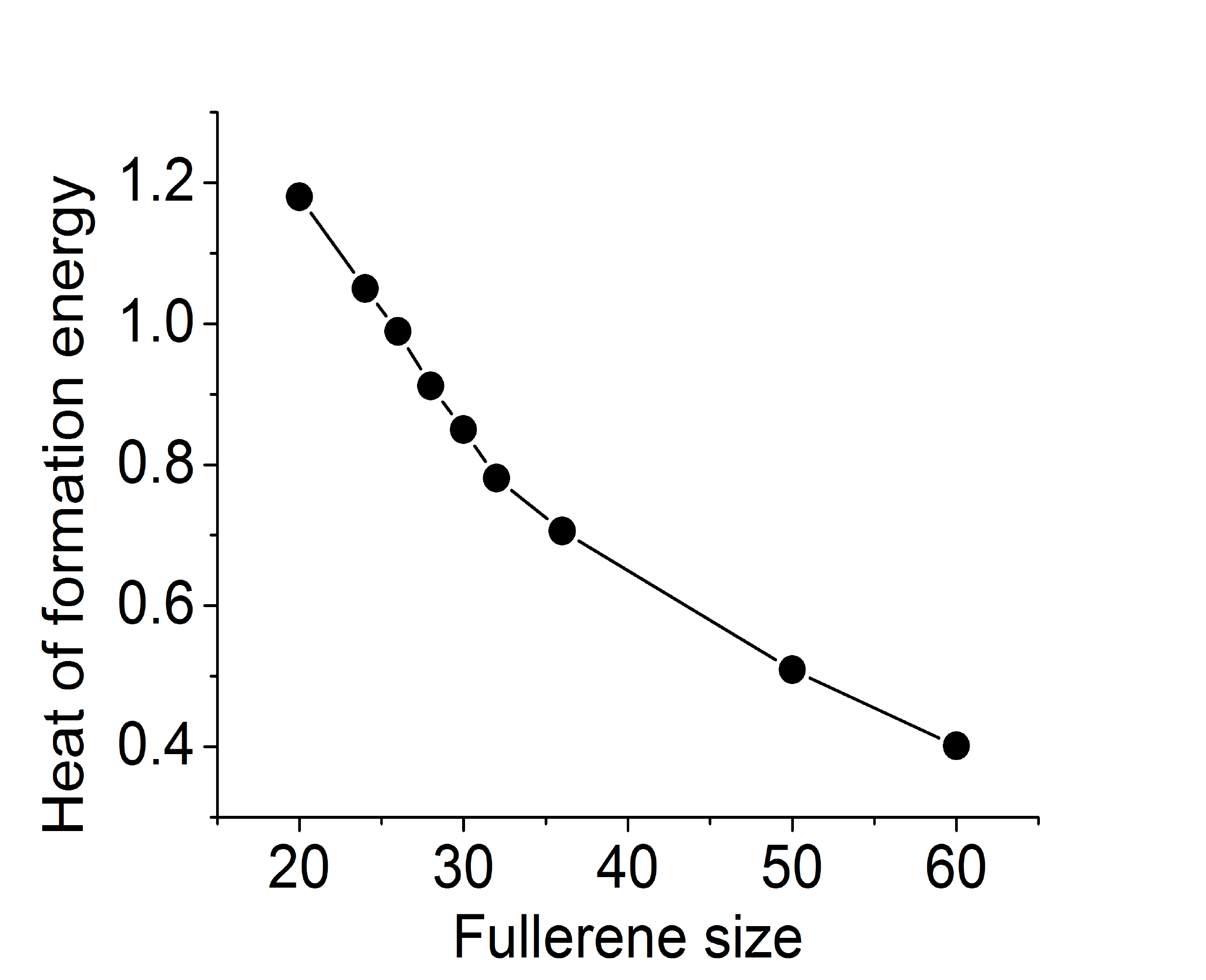}&
\includegraphics[width=0.45\textwidth]{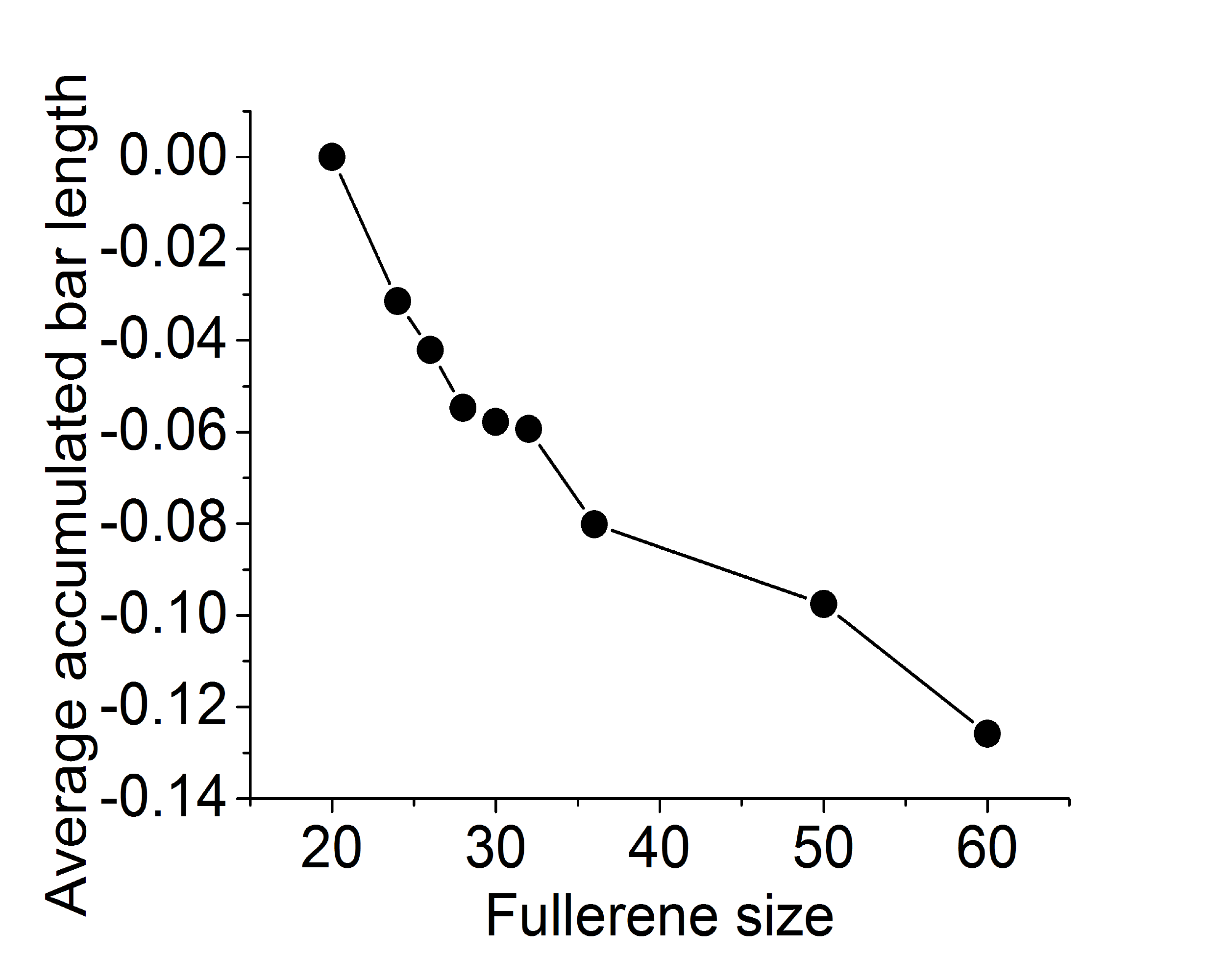}
\end{tabular}
\end{center}
\caption{Comparison between the heat of formation energies computed using a quantum theory  \cite{ZhangBL:1992a} (left chart) and  average accumulated bar length (right chart) for fullerenes. The units for the heat of formation energy and average accumulated bar length are eV/atom and  \AA/atom, respectively   Although the profile of average accumulated bar length of fullerenes does not perfectly match the fullerene energy profile, they bear a close resemblance in their basic characteristics.}
\label{fig:fullerenFitting}
\end{figure}

Zhang et al.  \cite{ZhangBL:1992a,ZhangBL:1992b} found that for small fullerene molecule series C$_{20}$ to C$_{70}$, their ground-state heat of formation energies gradually decrease with the increase of the number of atoms,  except for C$_{60}$ and C$_{70}$. The decreasing rate, however, slows down with the increase  of the number of atoms.   With data adopted from Ref. \cite{ZhangBL:1992a}, Fig.    \ref{fig:fullerenFitting} demonstrates this phenomenon. This type of behavior is also found in the total energy (STO-3G/SCF at MM3) per atom \cite{Murry:1994}, and in average binding energy of fullerene C$_{2n}$ which can be broken down to $n$ dimmers (C$_2$) \cite{ChangYF:2005}.

To understand this behavior, many theories  {have been} proposed. Zhang et al.~\cite{ZhangBL:1992b} postulate that the fullerene  stability is related to the ratio between the number of pentagons and the number of atoms for  a fullerene molecule. Higher percentage of pentagon structures results in relatively higher levels of the heat of formation. On the other hand, a rather straightforward isolated pentagon rule (IPR) states that the most stable fullerenes are those in which all the pentagons are isolated. The IPR explains why  C$_{60}$ and C$_{70}$ are relatively stable as both have only isolated pentagons. Raghavachari's neighbour index \cite{Raghavachari:1992} provides another approach to quantitatively characterize the relative stability. For example, in  C$_{60}$ of $I_n$ symmetry, all 12 pentagons have neighbour index 0, thus the $I_n$  C$_{60}$  structure is very stable.

In this work, we hypothesize that fullerene stability depends on the average number of hexagons per atom. The larger number of hexagons is in a given fullerene structure, the more stable it is. We utilize persistent homology to verify our hypothesis.  As stated in Section \ref{Sec:FullereneBarcodes}, there are two types of $\beta_2$ bars, namely, the one due to hexagon-structure-related holes and that due to the central void. Their contributions to the heat of formation energy are dramatically different.  Based on our hypothesis, we only need to include those $\beta_2$ bars that are due to  hexagon-structure-related holes in our calculation of the average accumulated bar length $B_2$. As depicted in the right chart of Fig.    \ref{fig:fullerenFitting}, the profile of the average accumulated bar length closely resembles that of the heat of formation energy. Instead of a linear decrease, both profiles exhibit a quick drop at first, then the decreasing rate slows down gradually. Although our predictions for C$_{30}$ and C$_{32}$ fullerenes do not match the corresponding energy profile precisely, which may be due to the fact that the data used in our calculation may  not be exactly the same ground-state data as those in the literature \cite{ZhangBL:1992b}, the basic characteristics and the relative relations in the energy profile are still well preserved. In fact, the jump at the C$_{60}$ fullerene is captured and stands out more obviously than the energy profile. This may be due to the fact that our method distinguishes not only pentagon and hexagon structures, but also the size differences within each of them.
We are not able to present the full set of energy data in Ref. \cite{ZhangBL:1992a} because we are limited by the availability of the ground-state structure data.

\begin{figure}
\begin{center}
\begin{tabular}{c}
\includegraphics[width=0.6\textwidth]{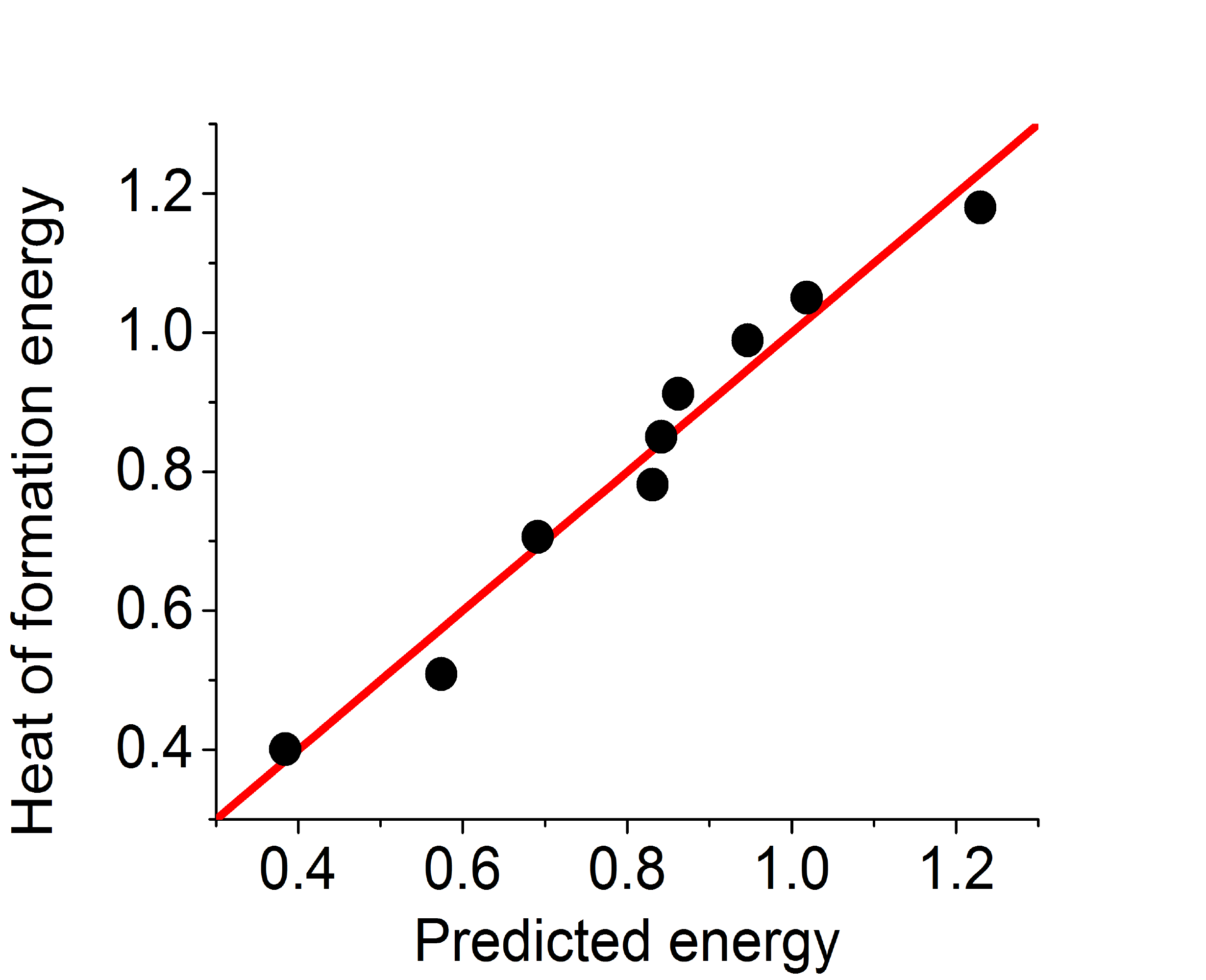}
\end{tabular}
\end{center}
\caption{The comparison between  {quantum mechanical simulation results}\cite{ZhangBL:1992a} and persistent homology prediction of the heat of formation energy (eV/atom). Only local $\beta_2$ bars that are due  to hexagon structures are included in our average accumulated bar length $B_2$.  The correlation coefficient from the least-squares fitting is near perfect ($C_c=0.985$).}
\label{fig:FitHeatFormation}
\end{figure}
 To quantitatively validate our prediction, the least squares method is employed to fit our prediction with the heat of formation energy, and a correlation coefficient is defined \cite{KLXia:2013d},
\begin{eqnarray}\label{correlation}
   C_c=\frac{\sum^N_{i=1}\left(B^e_i-\bar{B}^e \right)\left( B^t_i-\bar{B}^t \right)}
   { \left[\sum^N_{i=1}(B^e_i- \bar{B}^e)^2\sum^N_{i=1}(B^t_i-\bar{B}^t)^2\right]^{1/2}},
\end{eqnarray}
where $B^e_i$ represents the heat of formation energy of the $i$th fullerene molecule, and $B^e_t$ is our theoretical prediction. The parameter $\bar{B}^e$ and $\bar{B}^t$ are the corresponding mean values. The fitting result is demonstrated in Fig.    \ref{fig:FitHeatFormation}. The correlation coefficient is close to unity (0.985), which indicates the soundness of our model and the power of persistent homology for quantitative predictions.

\subsection{Total curvature energy analysis of fullerene isomers}

Having demonstrated the ability of persistent homology for the prediction of  the relative stability of fullerene molecules,  we further illustrate the effectiveness of persistent homology for analyzing the total curvature energies of fullerene isomers. Fullerene molecules  C$_N$  are well-known to admit various isomers \cite{Fowler:1996}, especially when the number ($N$) of atoms is large. In order to identify all of the possible isomers for a given $N$, many elegant mathematical algorithms have been proposed.  Coxeter's construction method \cite{Coxeter:1971,Fowler:1988} and  {the ring spiral method} \cite{Manolopoulos:1991} are two popular choices.  Before discussing the details of these two methods, we need to introduce the concept of fullerene dual. Mathematically, a dual means dimension-reversing dual. From Euler's polyhedron theorem, if a spherical polyhedron is composed of $n_{\rm vertice}$ vertices , $n_{\rm edge}$ edges and $n_{\rm face}$ faces, we have the relation $n_{\rm vertice}-n_{\rm edges}+n_{\rm face}=2$. Keeping the $n_{\rm edge}$ unchanged while swapping the other two counts, we have its dual, which has $n_{\rm vertice}$ faces and $n_{\rm face}$ vertices. For example, the cube and the octahedron form a dual pair, the dodecahedron and the icosahedron form another dual pair, and the tetrahedron is its self-dual. This duality is akin to the duality  between the Delaunay triangulation and the corresponding Voronoi diagram in computational geometry.

\begin{figure}
\begin{center}
\begin{tabular}{cc}
\includegraphics[width=0.45\textwidth]{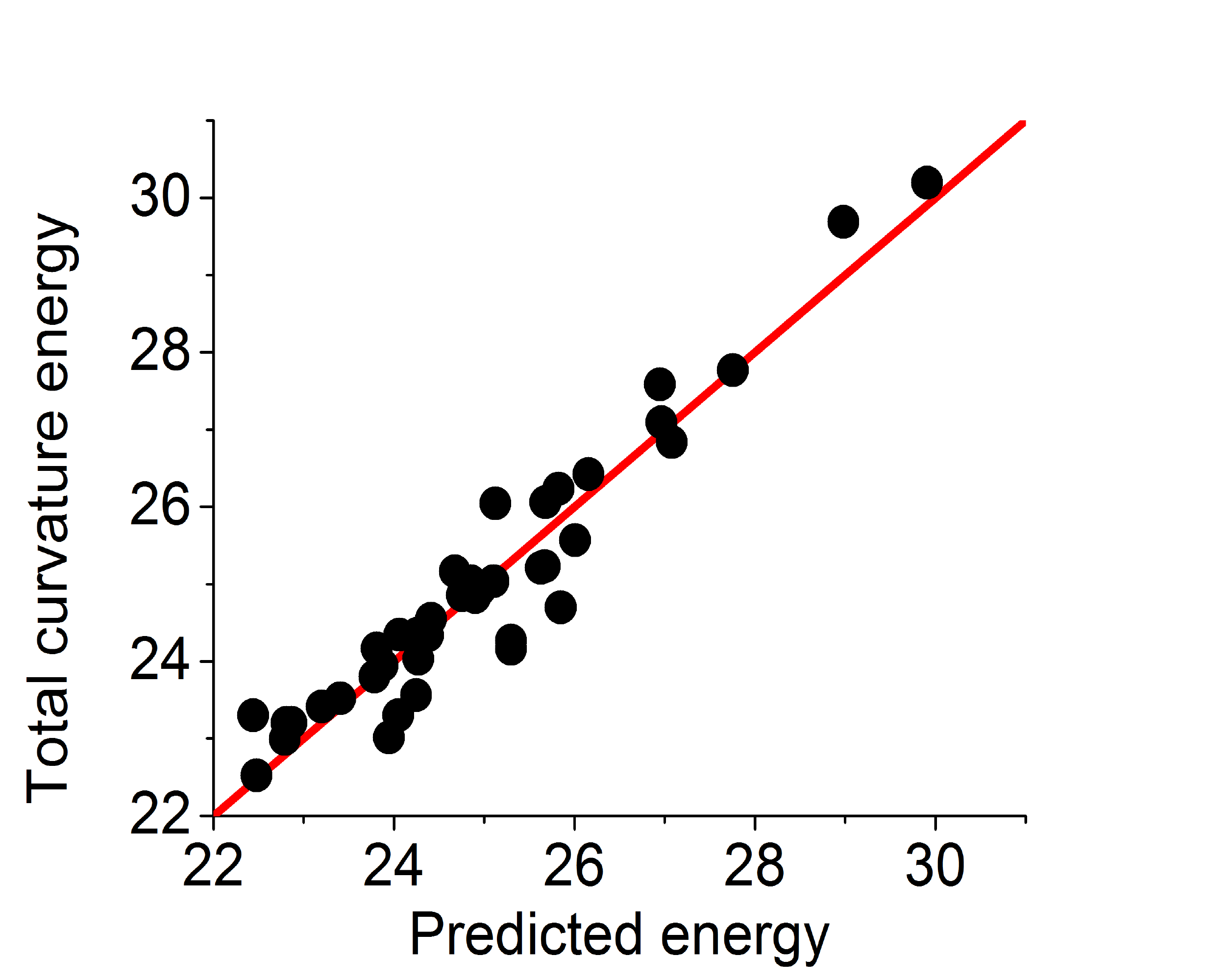}&
\includegraphics[width=0.45\textwidth]{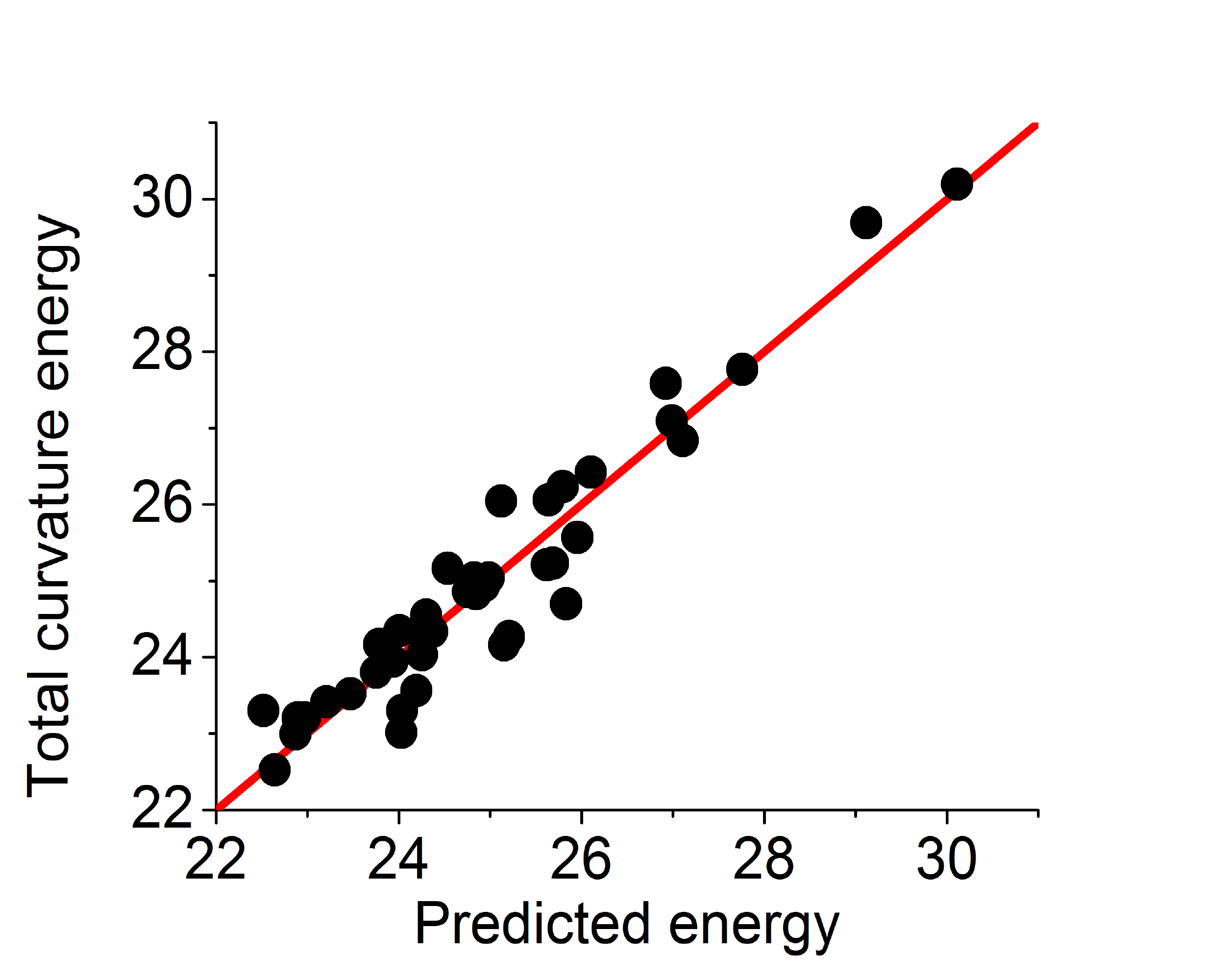}
\end{tabular}
\end{center}
\caption{Comparison between the distance filtration (left chart) and the correlation matrix filtration (right chart) in fullerene C$_{40}$ stability analysis. Fullerene C$_{40}$ has 40 isomers. Each of them has an associated total curvature energy (eV). We calculate our average accumulated bar lengths  from both distance filtration and the correlation matrix based filtration, and further fit them with  total curvature energies. The correlation coefficients for our fitting are 0.956 and 0.959, respectively.  It should be noticed that only the central void related $\beta_2$ bars (i.e., the long-lived bars) are considered. The exponential kernel is used in matrix filtration with parameter  $\eta=4$ and $\kappa=2$.}
\label{fig:IsomerC40}
\end{figure}

In fullerenes, each vertex is shared by three faces (each of them is either a pentagon or a hexagon). Therefore, fullerene dual can be represented as a triangulation of the topological sphere. Based on this fact, Coxeter is able to analyze the icosahedral triangulations of the sphere and predict the associated isomers.  This method, although mathematically rigorous, is difficult to implement for  {structures with low symmetry}, thus is inefficient in practical applications \cite{Fowler:1995}.
On the other hand, in the Schlegel diagram \cite{Schlegel:1883}, each fullerene structure can be projected into a planar graph made of pentagons and hexagons. The ring spiral method is developed based on the spiral conjecture \cite{Fowler:1995}, which states ``The surface of a fullerene polyhedron may be unwound in a continuous spiral strip of edge-sharing pentagons and hexagons such that each new face in the spiral after the second shares an edge with both (a) its immediate predecessor in the spiral and (b) the first face in the preceding spiral that still has an open edge.'' Basically, for  fullerenes of  $N$ atoms, one can list all possible spiral sequences of pentagons and hexagons, and then wind them up into fullerenes. If no conflict happens during the process, an isomer is generated. Otherwise, we neglect the spiral sequence. Table \ref{tab:Isomer} lists the numbers of isomers for different fullerenes \cite{Fowler:1995}, when enantiomers are regarded as equivalent \ref{tab:Isomer}. It is seen that the number of isomers increases dramatically as $N$ increases.
Total curvature energies of many fullerene isomers are available at the  \href{http://www.nanotube.msu.edu/fullerene}{carbon fullerene webpage}.

\begin{table}[htbp]
  \centering
\caption{Numbers of isomers  for small fullerenes. }
\begin{center}
\begin{tabular}{|c|c|c|c|c|c|c|c|c|c|c|c|c|}
 \hline
$N_{\rm atom}$ & 20  & 24  & 26   & 28 & 30 & 32 &34 & 36&38 &40 &50 &60\\
\hline
$N_{\rm isomer}$ & 1  & 1  & 1   & 2 & 3 & 6 &6 & 15 &17 &40 &271 &1812 \\
\hline
 \end{tabular}
  \label{tab:Isomer}
  \end{center}
\end{table}

In 1935, Hakon defined   sphericity as a measure of how spherical (round) an object is \cite{Hakon:1935}. By assuming particles having the same  volume but differing in surface areas, Hakon came up with  a sphericity function \cite{Hakon:1935},
\begin{eqnarray}
\Psi=\frac{\pi^{1/3} (6V_p)^{2/3}}{A_p},
\end{eqnarray}
 where  $V_p$  and $A_p$ are the volume and  the  surface area of the particle. Obviously, a sphere has  sphericity 1, while the sphericity of non-spherical particles is less than 1. Let us assume that fullerene isomers have the same surface area as the perfect sphere $A_p=4\pi R^2$, we define a sphericity measure as
\begin{eqnarray}
\Psi_c=\frac{V_p}{V_s}
      = \frac{6\pi^{1/2} V_p}{A_p^{3/2}},
\end{eqnarray}
where $V_s$ is the volume of a sphere with radius $R$.
By the isoperimetric inequality,  among all simple closed surfaces with given surface area $A_p$, the sphere encloses a region of maximal volume.
 Thus,  the sphericity of non-spherical fullerene isomers is less than 1.  Consequently, in a distance based filtration process, the smaller
sphericity a fullerene isomer is,  the  shorter its global $\beta_2$ bar will be.

On fullerene surface, the local curvature characterizes the bond bending away from the plane structure required by the sp$^2$ hybrid orbitals \cite{Holec:2010}. Therefore, the relation between fullerene curvature and  stability  can be established  and confirmed by using {\it ab initio} density functional calculations \cite{Guan:2014}. However, such an analysis favors fullerenes with infinitely many atoms. Let us keep the assumption that  for a given fullerene C$_N$, all its isomers have the same surface area. We also assume that the most stable fullerene isomer C$_N$ is the one that has a near perfect spherical shape. Therefore, each fullerene isomer is  subject to a (relative) total curvature energy $E_c$  per unit area due to  its accumulated  deviations from a perfect sphere,
\begin{eqnarray}\label{CuravtureE}
E_c&=&\int_\Gamma \mu\left[(\kappa_1-\kappa_0)^2 + (\kappa_2-\kappa_0)^2 \right] dS\\
&=& \int_\Gamma 2\mu \left[\frac{1}{2}(2{\bf H}-\kappa_0)^2 + {\bf K}  \right] dS,
\end{eqnarray}
where $\Gamma$ is the surface, $\mu$ is bending rigidity,   $\kappa_1$ and  $\kappa_2$ are the two principal curvatures, and $\kappa_0=1/R$ is the constant curvature of the sphere with radius $R$.  Here, ${\bf H}$ and ${\bf K}$ are the mean and Gaussian curvature of the fullerene surface, respectively. Therefore, a fullerene isomer with a smaller sphericity will have a higher total curvature energy. Based on the above discussions, we  establish the inverse correlation between fullerene isomer global $\beta_2$ bar lengths and   fullerene isomer total curvature energies.

Obviously, the present fullerene curvature energy (\ref{CuravtureE}) is a special case of  the Helfrich energy functional for elasticity of cell membranes \cite{Helfrich:1973}
\begin{eqnarray}
E_c=\int_\Gamma   \left[ \frac{1 }{2}{\cal K}_C(2{\bf H}-C_0)^2 + {\cal K}_G {\bf K}  \right] dS,
\end{eqnarray}
where, $C_0$ is the spontaneous curvature, and  ${\cal K}_C$ and ${\cal K}_G$ are the bending modulus and Gaussian saddle-splay modulus,
respectively.  The Gauss\ - Bonnet theorem states that for a compact two-dimensional Riemannian manifold without boundary, the surface integral of the Gaussian curvature is $2\pi \chi$, where $\chi$ is the Euler characteristic. Therefore, the curvature energy admits a jump whenever there is a change in topology which leads to a change in the Euler characteristic. A problem with this discontinuity in the curvature energy is that  the topological change may be induced by an  infinitesimal change in the geometry associated with just an infinitesimal physical energy, which implies that the Gaussian   curvature energy functional is unphysical. Similarly, Hadwiger type of  {energy} functionals, which make use of a linear combination of the surface area, surfaced enclosed volume,  and surface integral of mean curvature and surface integral of Gaussian curvature \cite{Hadwiger:1975}, may be unphysical as well for systems involving topological changes.  However, this is not a problem for differential geometry based multiscale models which utilize only  {surface} area and  {surface} enclosed volume terms \cite{Wei:2009,Wei:2012,Wei:2013,ZhanChen:2010a},  {as we employ the Eulerian representation and the proposed generalized mean curvature terms but not Gaussian curvature terms.} Moreover, in the present model for fullerene isomers, there is no topological change.

\begin{figure}
\begin{center}
\begin{tabular}{cc}
\includegraphics[width=0.45\textwidth]{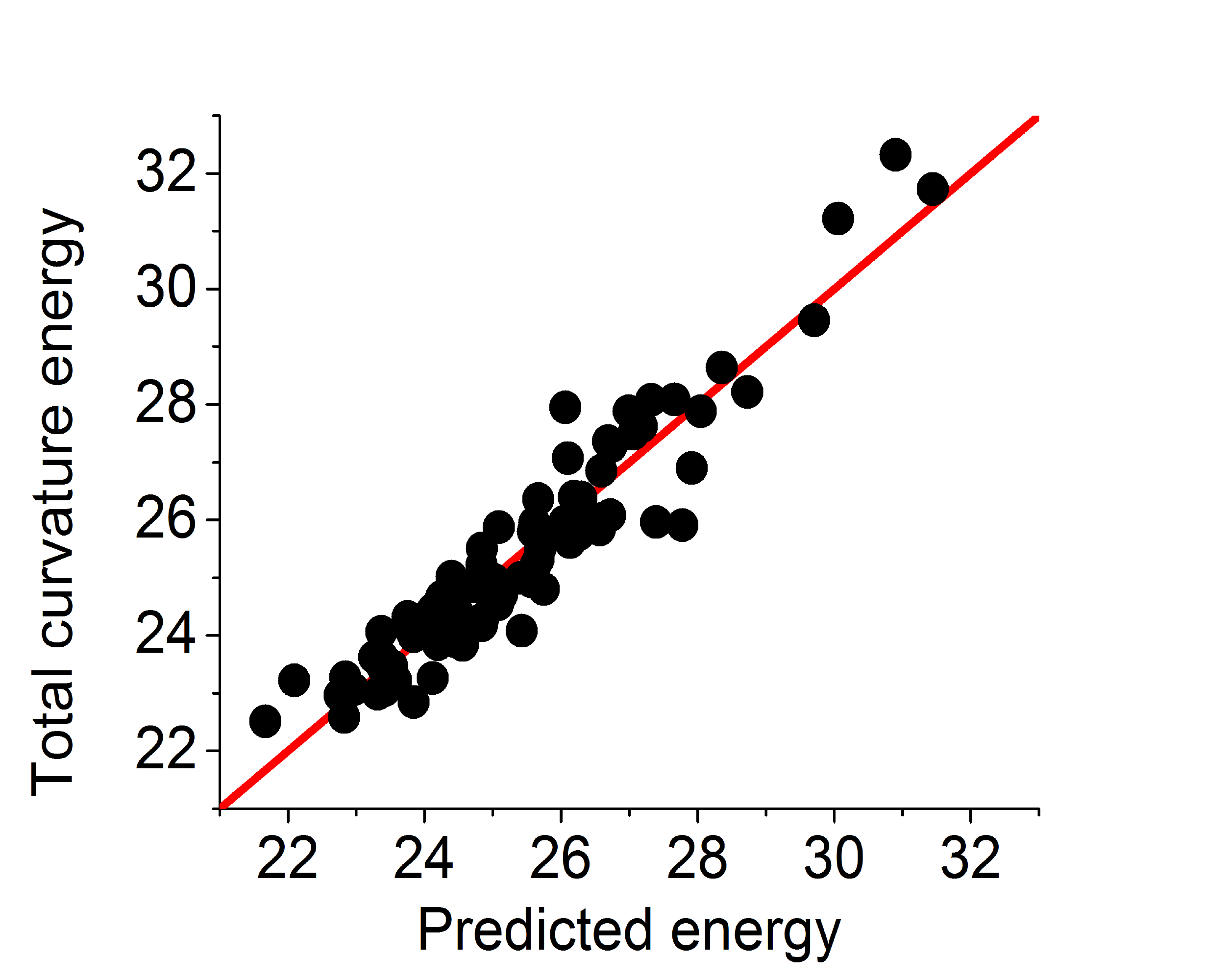}&
\includegraphics[width=0.45\textwidth]{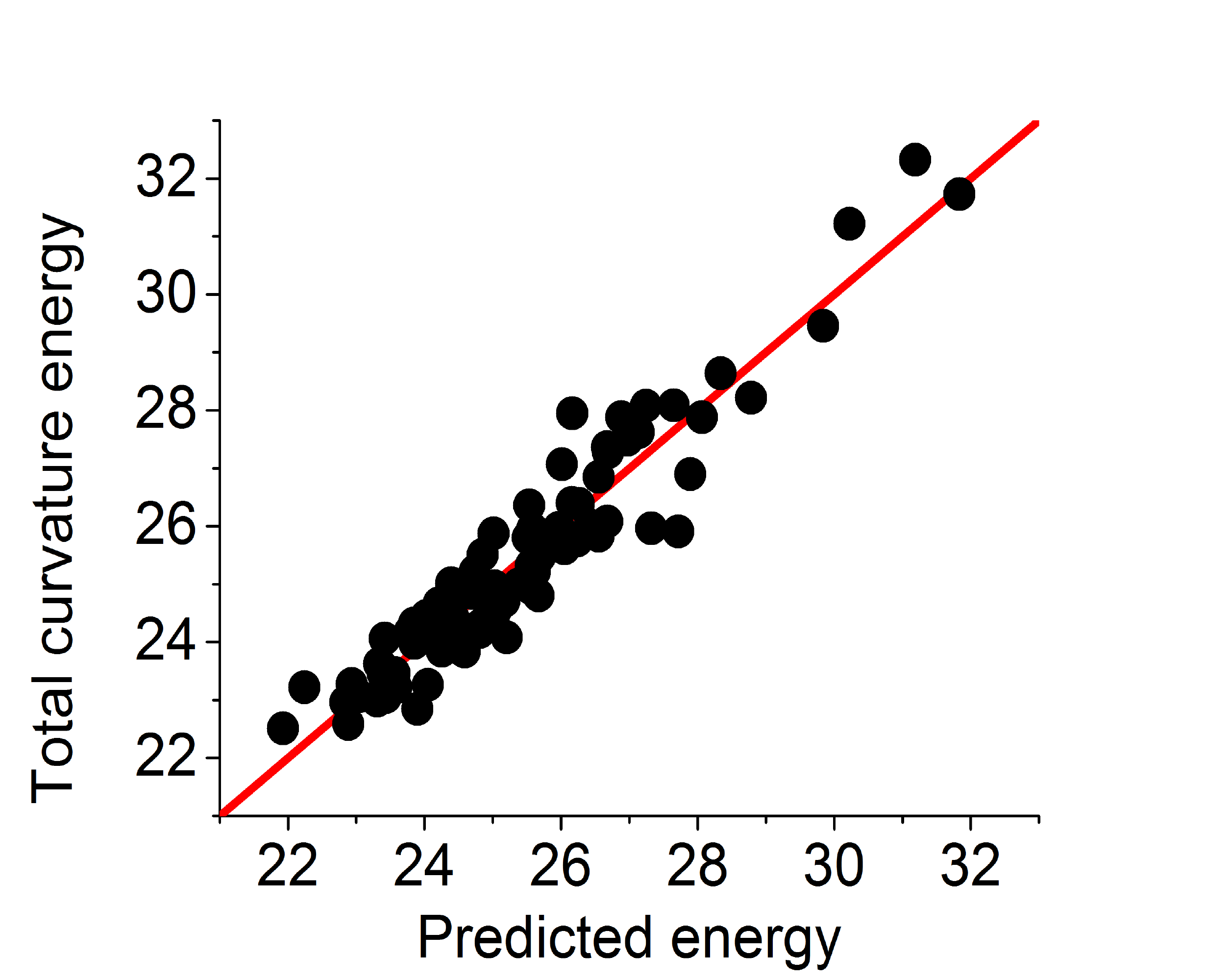}
\end{tabular}
\end{center}
\caption{Further validation of our method with  89 isomers for fullerene C$_{44}$. The correlation coefficients for distance filtration (left chart) and correlation matrix based filtration (right chart) are 0.948 and 0.952, respectively. In the latter method, the exponential kernel is used with parameter $\eta=4$ and $\kappa=2$.
}
\label{fig:IsomerC44}
\end{figure}

To verify our assumptions, we consider a family of isomers for fullerene C$_{40}$. It has a total of 40 isomers.  We compute the global $\beta_2$ bar lengths of all isomers  by Euclidean distance filtration and fit their values with their total curvature energies   with a negative sign.    Figure \ref{fig:IsomerC40} (right chart) shows an excellent correlation between the fullerene total curvature energies and our persistent homology based predictions. The correlation coefficient is 0.956, which indicates that the proposed  persistent homology analysis of non-sphericity and  our assumption of a constant surface area  for all fullerene isomers are sound. In reality, fullerene isomers may not have an exactly constant surface area because some distorted bonds may have a longer bond length. However, the high    correlation coefficient  found in our persistent homology analysis implies that either the average   bond lengths for all isomers are similar or the  error due to non-constant surface area is offset by other errors.

To further validate our persistent homology based method for the prediction of fullerene total curvature energies, we consider a case with significantly more isomers, namely,   fullerene  C$_{44}$, which has 89 isomers. In this study, we have again found an  excellent correlation between the fullerene total curvature energies and our persistent homology based predictions as depicted in the right chart of Fig.    \ref{fig:IsomerC44}.  The correlation coefficient for this case is 0.948. In fact, we have checked more fullerene isomer systems and obtained similar  predictions.

\begin{figure}
\begin{center}
\begin{tabular}{c}
\includegraphics[width=0.7\textwidth]{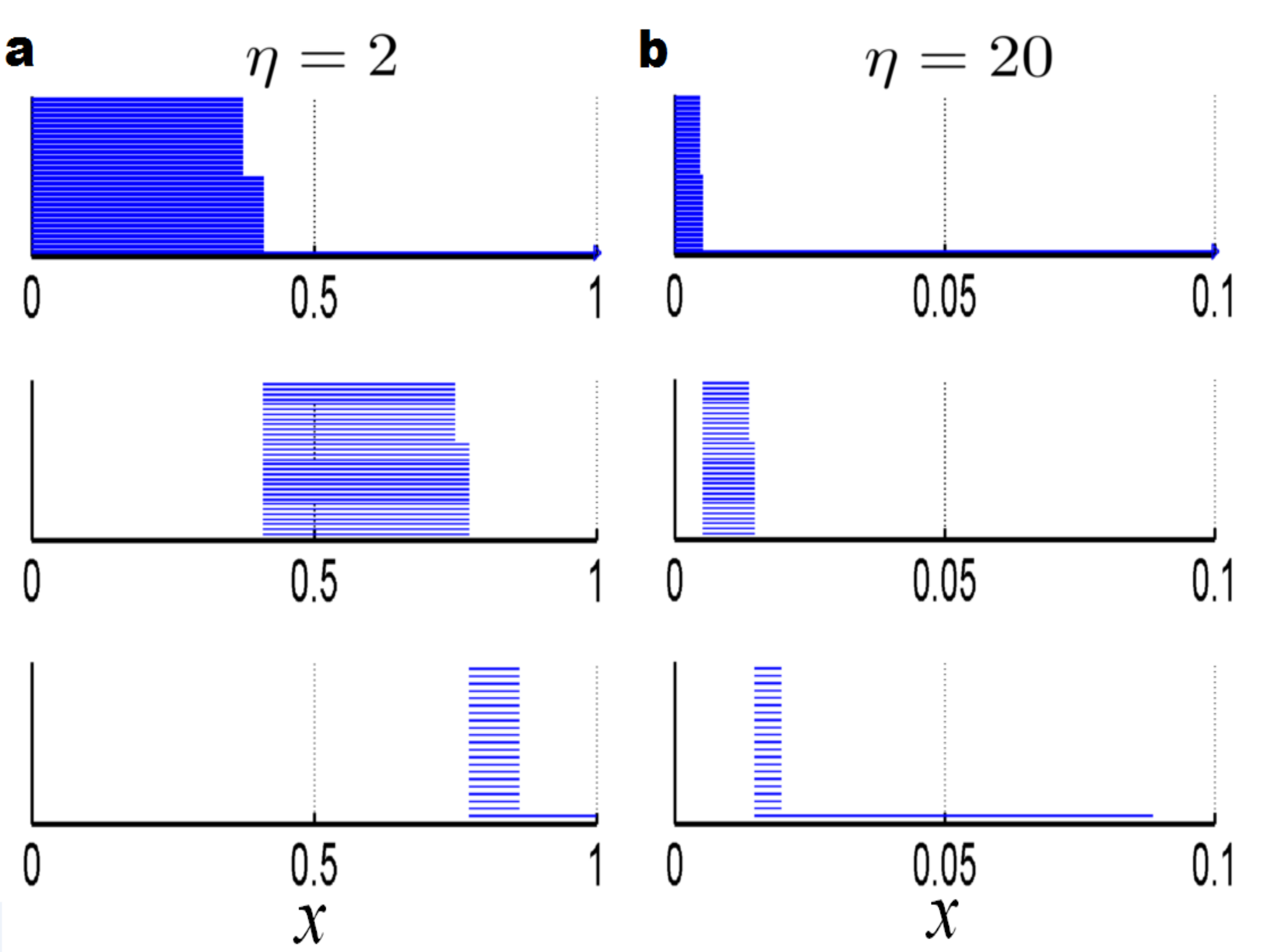}
\end{tabular}
\end{center}
\caption{Illustration of the persistent barcodes generated by using  correlation matrix based filtrations with different characteristic distances. The exponential kernel model with power $\kappa=2$ is used. The characteristic distances in the left and right charts are respectively $\eta=2$ and $\eta=20$. }
\label{fig:Sigma_barcodes}
\end{figure}

Finally, we explore the  utility of our correlation matrix based filtration process for analysis of fullerene total curvature energies.  In place of  Euclidean distance based filtration, the correlation matrix based filtration is employed. To demonstrate the basic principle, Eq.~(\ref{eq:FiltrationMatrix}) with the generalized exponential kernel in Eq. (\ref{eq:ExpKernel}) is used in the filtration. We assume $w_{ij}=1$ as fullerene molecules have only carbon atoms.  To understand the correlation matrix based filtration method, the fullerene C$_{60}$ is employed again. We fixed the power $\kappa=2$, and adjust the value of characteristic distance $\eta$. Figure \ref{fig:Sigma_barcodes} gives the calculated barcodes with $\eta=2$ and $\eta=20$. It can be seen that these barcodes share a great similarity with the Euclidean distance based filtration results depicted in the right chart of Figure \ref{fig:C20C60Barcode}. All of topological features, namely,  two kinds of bonds in $\beta_0$, the pentagonal rings and the hexagonal rings in $\beta_1$, and also the hexagonal cavities and the central void in $\beta_2$ are clearly demonstrated.  However, it should be noticed that, unlike the distance based filtration, the matrix filtration  does not generate linear Euclidean distance relations. However,  relative correspondences within the structure are kept. For instances, in $\beta_2$ bars, the bar length ratio between the central void part and the hexagonal hole part in Fig.    \ref{fig:Sigma_barcodes} is drastically different from its counterpart in Fig.    \ref{fig:C20C60Barcode}. From our previous experience in flexibility and rigidity analysis \cite{KLXia:2013d,KLXia:2013f,KLXia:2014b}, these rescaled distance relations have a great potential in capturing the essential physical properties, such as, flexibility, rigidity, stability, and compressibility  of the underlying system.

Similarly, the global $\beta_2$ bar lengths obtained  from   the correlation matrix based filtration are utilized to fit with the total curvature energies of fullerene isomers.  The correlation coefficients for the correlation distance matrix filtration are  0.959 and 0.952, respectively for C$_{40}$ and C$_{44}$ fullerene isomers.  The corresponding results are demonstrated in the right charts of Figs. \ref{fig:IsomerC40} and \ref{fig:IsomerC44}, respectively. It can be seen that the correlation matrix filtration is able to capture the essential stability behavior of fullerene isomers. In fact,  results from correlation matrix based filtrations  are slightly better than those of Euclidean distance based filtrations. In correlation matrix based filtrations, the generalized exponential kernel is used with parameter $\eta=4$ and $\kappa=2$. These parameters are chosen based on our previous flexibility and rigidity analysis of protein molecules. Overall, best prediction is obtained when the characteristic distance is about 2 to 3 times of the bond length and power index $\kappa$ is around 2 to 3. Fine tuning of the parameters for each single case  may yield even better result. However, this aspect is beyond the scope of  the present work.

\section{Conclusion}\label{Sec:Conclusion}

Persistent homology is an efficient tool for the qualitative analysis of topological features that last over scales. In the present work, for the first time, persistent homology  is introduced for the quantitative prediction of fullerene energy and stability. We briefly review the principal concepts and algorithms in persistent homology, including simplex, simplicial complex, chain, filtration, persistence, and paring algorithms. Euler characteristics analysis is employed to decipher the barcode representation of fullerene C$_{20}$ and C$_{60}$. A thorough understanding of fullerene barcode origins enables us to construct physical models based on local and/or global topological invariants and their accumulated persistent lengths. By means of an average accumulated bar length of the second Betti number that corresponds to fullerene hexagons,  we are able to  accurately predict the relative energies of a series of small fullerenes. To analyze the total curvature energies of fullerene isomers, we propose to use sphericity to quantify the non-spherical fullerene isomers  and correlate the sphericity with fullerene isomer total curvature energies, which are defined as a special case of the Helfrich energy functional for elasticity. Topologically, the sphericity of a fullerene isomer is measured by its global 2nd homology bar length in the barcode, which in turn gives rise to the prediction of fullerene isomer total curvature energies.   We demonstrate an excellent agreement between  total curvature energies and our persistent homology predictions for the isomers of fullerene C$_{4}$ and C$_{44}$. Finally, a new filtration based on the correlation  matrix of the flexibility and rigidity index is proposed and found to provide even more accurate  predictions of fullerene  isomer total curvature energies.

\vspace{1cm}

\noindent \textbf{Acknowledgments}\\
\noindent  This work was supported in part by NSF grants  IIS-0953096, IIS-1302285 and  DMS-1160352,  NIH grant  R01GM-090208 and  MSU   Center for Mathematical Molecular Biosciences initiative.
GWW acknowledges the Mathematical Biosciences Institute for hosting valuable workshops. KLX thanks Bao Wang for useful discussions.

\vspace{1cm}

\footnotesize 



\end{document}